\documentclass[journal]{IEEEtran}
\usepackage{amsmath}
\usepackage{relsize}
\usepackage{bm}
\usepackage{amsfonts}
\usepackage{amssymb}
\usepackage{amscd}
\usepackage{graphicx}
\usepackage{epsfig}
\usepackage{subfigure}
\usepackage{newlfont}
\usepackage{cite}
\usepackage{enumitem}
\usepackage{amsthm}
\usepackage{verbatim}
\usepackage{xcolor}

\IEEEoverridecommandlockouts
\newcommand{\setxysizeo}{\epsfxsize=2.7in}
\newcommand{\pullUp}{\vskip -0.0cm}

\begin{document}


\title{Orthogonal Delay Scale Space Modulation: A New Technique for Wideband Time-Varying Channels} 

\author{\IEEEauthorblockN{Arunkumar K.~P. and Chandra R. Murthy}
\thanks{Arunkumar K.~P. is with the Naval Physical Oceanographic Laboratory, Kochi 682 021, India (e-mail: arunkumar@npol.drdo.in). This work was carried out as part of the PhD program at Indian Institute of Science, Bangalore. Chandra R. Murthy is with the Dept. of ECE, Indian Institute of Science, Bangalore 560 012, India (e-mail: cmurthy@iisc.ac.in).}
}


\maketitle

\begin{abstract}
	  Orthogonal Time Frequency Space (OTFS) modulation is a recently proposed scheme for time-varying narrowband channels in terrestrial radio-frequency communications.  Underwater acoustic (UWA) and ultra-wideband (UWB) communication systems, on the other hand, confront wideband time-varying channels. Unlike narrowband channels, for which time contractions or dilations due to \textcolor{black}{Doppler effect} can be approximated by frequency-shifts, the Doppler effect in wideband channels \textcolor{black}{results} in frequency-dependent non-uniform shift of signal frequencies across the band. In this paper, we develop an OTFS-like modulation scheme -- Orthogonal Delay Scale Space (ODSS) modulation -- for handling wideband time-varying channels. We derive the ODSS transmission and reception schemes from first principles. In the process, we introduce the notion of $\omega$-\emph{convolution} in the \emph{delay-scale space} that parallels the \emph{twisted convolution} used in the \emph{time-frequency space}.
	  The preprocessing 2D transformation from the Fourier-Mellin domain to the delay-scale space in ODSS, which plays the role of inverse symplectic Fourier transform (ISFFT) in OTFS, improves the bit error rate performance  compared to OTFS and Orthogonal Frequency Division Multiplexing (OFDM) in wideband time-varying channels. Furthermore, since the channel matrix is rendered near-diagonal, ODSS retains the advantage of OFDM in terms of its low-complexity receiver structure.
\end{abstract}
\begin{keywords}
	Underwater acoustic/ultra-wideband communications, wideband time-varying channels, Mellin transform.
\end{keywords}

\section{Introduction} \label{sec:intro}

Orthogonal frequency division multiplexing (OFDM) is a spectrally efficient scheme for communication over frequency-selective channels. The scheme is particularly attractive in practice because a low complexity receiver side processing based on \textcolor{black}{a subcarrier-by-subcarrier equalization} recovers the data symbols in a delay spread channel. Receivers based on \textcolor{black}{subcarrier-by-subcarrier equalizer}, however, fail in a time-varying channel, resulting in severe degradation of communication performance.  Orthogonal Time Frequency Space (OTFS) modulation is a recently proposed technique for use in frequency-selective and Doppler spread narrowband channels \cite{HadaniMar2017, Hadani2018Orthogonal, Hadani2018otfs, HadaniIMS2017, RavitejaJune2019, Murali2018}. OFDM and OTFS techniques do not perform well in doubly-spread wideband channels where the effect of Doppler is to cause a time-scaling in the received waveform. 

In high mobility \emph{narrowband} channels, characterized by both delay spread (due to multipath) and Doppler spread (due to time variations and/or mobility), the OTFS scheme achieves a near constant gain channel for each subcarrier. The scheme employs special transformations at the transmitter to mount the information symbols on the carrier waveform, and corresponding inverse transformations to recover those symbols at the receiver. In wideband channels where the Doppler manifests as a time-scaling of the received waveform, however, the channel is no longer flat-fading across the OTFS subcarriers. \textcolor{black}{The extent of detriment caused by a seemingly small time-scale factor, such as $\alpha = 1.001$, in wideband channels, if not handled well, is akin to the consequences narrated in \cite{ObergJames2004}. Simply increasing the transmission power does not help in recovering the data from a waveform affected by Doppler. In \cite{ObergJames2004}, the navigation experts had to alter the descent trajectory of Huygens (transmitting probe) such that its descent to Titan (Saturn's moon) is almost perpendicular to the line joining Cassini (receiver). This contains the radial component of probe's velocity, thereby mitigating the stretching of the communication waveform due to Doppler. Avoiding Doppler is not possible in all practical situations. Our goal in this work is to develop a modulation scheme, suitable for \emph{wideband} time-varying channels, with a demodulation counterpart that can be implemented as a low complexity receiver.}

Wideband and ultra-wideband channel models abound in the literature on wireless communications \cite{Franz, Rickard, Balan, Molisch2014, Molisch, Schuster, Benedetto, Nikookar, Giannakis, Haimovich, Margetts, Tse}. High mobility wireless channels are of topical interest in broadband high speed radio-frequency communications \cite{Fan, He, Wu2016ASO, Sammna}. Wideband time-varying channel models are commonly used in underwater acoustic communications also \cite{Davies, Stojanovic96, Stojanovic2013, Papandreou, PapandreouConf, Preisig, Roudsari2017ChannelMF, Zakharov, Fengzhong}. Several studies have considered the problem of communicating data over wideband doubly-spread channels \cite{Eggen, Dhanoa, Sadia2008, Freitag2006, Berger2008, Berger2010SStoCS, Berger2011ProgICIEq, Schlegel, Shengli2007, Yerramalli2012PartialFFT, SparseComparisonHuang, Li2008Nonuniform}. \textcolor{black}{Two approaches have been explored in the literature. One approach is to estimate and compensate for the effect of Doppler in the received signal, equalize the effect of delay spread, and decode the data symbols~\cite{Freitag1997, Shengli2007, Li2008Nonuniform, Berger2008}. Despite such compensation, residual Doppler in the processed signal affects the communication performance in a multipath environment, since different paths are associated with different amounts of Doppler. The residual Doppler causes inter carrier interference (ICI) in multi-carrier communication systems. The second approach uses computationally expensive receivers that account for the ICI in data detection~\cite{Berger2010SStoCS, Yerramalli2012PartialFFT, Berger2011ProgICIEq, SparseComparisonHuang, Schlegel}. The recently proposed OTFS scheme is specifically developed to handle doubly-spread narrowband channels, and offer high performance, but at the cost of a more sophisticated message-passing based receiver architecture. To the best of our knowledge, no such scheme has been proposed in the literature for the case of wideband doubly-spread channels.} 

In this paper, we systematically develop the processing blocks of a new modulation scheme -- Orthogonal Delay Scale Space (ODSS) modulation. Specifically, inspired by the development of OTFS in \cite{Hadani2018otfs, Hadani2018Orthogonal} for narrowband time-varying channels, we parallel its development by identifying the transformations necessary to handle the time-scaling effect of a wideband time-varying channel. Our contributions in this work are:
\begin{enumerate}
	\item  We derive the ODSS transmission and reception schemes from first principles. We identify the  modulation and demodulation operations that are required for handling the time-scaling effect of wideband time-varying channels. In particular, we introduce a preprocessing 2D transformation from the Fourier-Mellin domain to the delay-scale space in ODSS. This transform plays the role of inverse symplectic Fourier transform (ISFFT) in OTFS. In contrast to the constant spectral width of the OFDM and OTFS subcarriers, the  subcarriers of ODSS have a spectral width that is proportional to the subcarrier frequency, which makes ODSS suitable for time-scale spread channels.
	\item \textcolor{black}{We introduce the notions of $\omega$-\emph{convolution} and \emph{robust bi-orthogonality} in the \emph{delay-scale space}. These parallel the notions of twisted convolution and robust bi-orthogonality, respectively, in the \emph{time-frequency space} in the OTFS scheme.}
	\item We analytically derive conditions on the parameter values of the ODSS scheme that results in an ICI-free symbol reception at the receiver. As a consequence, the ODSS receiver is a low complexity processor that uses a subcarrier-by-subcarrier equalizer as in the case of OFDM over a time-invariant inter symbol interference channel.
	\item We compare the performance of the ODSS scheme with OFDM and OTFS that uses a low complexity receiver based on \textcolor{black}{subcarrier-by-subcarrier equalization}. The ODSS receiver registers more than $100$ fold reduction in the bit error rate (BER) compared to the OFDM and OTFS receivers employing \textcolor{black}{subcarrier-by-subcarrier equalizers} at an SNR of $24$~dB.
\end{enumerate}

Wideband doubly-spread (also known as  multi-scale multi-lag) channels are found in underwater acoustic  (UWA) and ultra wideband (UWB) radio communications. 
Low complexity \textcolor{black}{subcarrier-by-subcarrier equalizers} of standard OFDM and OTFS receivers suffer from performance impairment whose severity increases with Doppler spread. The ODSS scheme using \textcolor{black}{subcarrier-by-subcarrier equalizer} based receiver, developed in this paper, is therefore promising in such channels, particularly in applications that require a low complexity receiver.

We briefly describe the narrowband and wideband time-varying channel models in Section~\ref{sec:DSchannel}. In Section~\ref{sec:OTFS}, we review the OTFS scheme devised for narrowband time-varying channels. We present the Mellin transformation and its properties in Section~\ref{sec:MellinX}, which forms a part of the preprocessing transformation in the ODSS scheme. In Section~\ref{sec:ODSS}, we develop the ODSS modulation and demodulation schemes and derive conditions on its parameters to make its \textcolor{black}{output ICI-free}. Section~\ref{sec:pulseShapeFn} discusses practical aspects of choosing transmit and receive filters that result in nearly ICI-free ODSS outputs.
Finally, through numerical simulations, we investigate the performance of ODSS in Section~\ref{sec:numResults} and conclude in Section~\ref{sec:conclusions}.

\section{Doubly Spread Channel Models} \label{sec:DSchannel}

A transmitted signal undergoes three changes when passing through a delay-scale propagation channel: (a) amplitude change due to path loss and fading, (b) delay, $\tau$, corresponding to the length of the path traversed, and (c) time-scaling by a factor, $\alpha = \frac{c-v}{c+v}$, due to Doppler effect, where $v$ is the velocity of a scatterer and $c$ is the speed of the wave in the propagation medium. Multiple propagation paths can result in a continuum of delay and scale parameters, i.e., $\tau \in [\tau_l, \tau_h]$ and $\alpha \in [\alpha_l, \alpha_h]$. Such a \emph{doubly-spread} propagation channel is characterized by the wideband spreading function, $h(\tau, \alpha)$, that corresponds to the amplitude gain of the time-scaled and delayed copy of the transmitted signal reaching the receiver along a reflected path. The received signal is a superposition of the amplitude-scaled, time-scaled, and delayed versions of the transmitted signal, $s(t)$, given by \cite{Franz}
\begin{equation} \label{eqn:wbDelScaleChan}
r_s(t) = \iint  h(\tau, \alpha) \sqrt{\alpha} s\left( \alpha (t-\tau) \right)d\tau d\alpha,
\end{equation}
where we have omitted the limits of integration, which we do throughout this paper, for notational brevity. \textcolor{black}{Note that the scaling by $\sqrt{\alpha}$ in the integrand above preserves the energy of the time-scaled copy of the transmitted signal, since $\int \left|  \sqrt{\alpha} s(\alpha t) \right|^2 dt = \int \left| s(t) \right|^2 dt$.}

The delay-scale channel representation used in \eqref{eqn:wbDelScaleChan} is called the \emph{wideband channel model}. Modeling the Doppler effect by approximating the time-scale by a frequency shift, i.e., using $\nu \approx (\alpha-1) f_c$, where $f_c$ is the center frequency of the signal band, leads to the narrowband model. The narrowband approximation holds if two conditions are met \cite{Papandreou}:
\begin{description}
	\item{(A-1)} Signal has a small fractional bandwidth: $B/f_c \ll 1$, where $B$ is the signal bandwidth, and
	\item{(A-2)} The receiver moves slowly relative to the transmitter, such that its position does not change significantly compared to the positional resolution of the signal: $v \ll \frac{c}{2 BT}$, where $T$ is the signal duration.\footnote{In the case of OFDM, $T$ is the duration of an OFDM symbol.}
\end{description}

\textcolor{black}{Under the assumptions above, the signal reflected by a scatterer moving with velocity $v$ and arriving along a path of delay, $\tau \in [\tau_l, \tau_h]$, is given by $s_{\tau,\nu} = s(t-\tau) e^{j2\pi \nu (t-\tau)}$, where $j = \sqrt{-1}$, and $\nu = (\alpha - 1)f_c \approx -\frac{2v}{c} f_c \in [\nu_l, \nu_h]$ is the frequency shift due to the Doppler effect \cite{Sibul1994}.  Making the change of variables $\alpha \rightarrow \nu\colon \alpha = 1 +\nu/f_c$ in \eqref{eqn:wbDelScaleChan}, noting that $\sqrt{\alpha} s\left( \alpha (t-\tau) \right) \approx s_{\tau,\nu} = s(t-\tau) e^{j2\pi \nu (t-\tau)}$ under narrowband assumptions, and defining the narrowband channel spreading function to be $h_{\tau,\nu}(\tau, \nu) \triangleq \frac{1}{f_c} h(\tau, \alpha = 1 + \nu/fc)$, we get the following expression for the received signal at the output of a narrowband channel \cite{Franz, Hadani2018Orthogonal}}:
\begin{equation} \label{eqn:nbDelDoplChan}
r_s(t) = \iint h(\tau, \nu)  s\left( t-\tau\right)e^{j2\pi \nu \left( t - \tau \right)} d\tau d\nu,
\end{equation}
\textcolor{black}{where we drop the subscripts in $h(\tau, \nu)$ for notational brevity.} 

\textcolor{black}{\emph{Remark 1:} Violation of either (A-1) or (A-2) would require one to model the channel as wideband. For example, UWA communications over a frequency band of $10$-$20$ kHz has a fractional bandwidth of $B/f_c = 0.67$  ($B/f_c> 0.25$ is considered high~\cite{Papandreou}). The low speed of sound in water also results in a high $v/c$ ratio ($10^{-3} - 10^{-1}$) that violates (A-2). Similarly, large time-bandwidth product $(10^{5} - 10^{6})$ UWB radio-frequency communications may violate (A-2) \cite{Margetts, Papandreou}. For example, a radio-frequency UWB communication with $BT = 10^6$, from a high-speed train traveling at $v = 270$~km/h, clearly violates (A-2).} 

\textcolor{black}{\emph{Remark 2:} The notion of \emph{wideband channel} we use here is different from the definition of a \emph{wideband system} used in the communications literature. A wideband system is one for which the signaling (messaging) bandwidth significantly exceeds the coherence bandwidth $(\propto \frac{1}{\tau_h - \tau_l})$ of the channel. The notion of wideband channel we use here is related to the frequency-dependent effect of the Doppler whereas the latter definition is related to the channel delay-spread.}

In the next section, we review the development of OTFS communication for the narrowband channel model as prelude to the development of ODSS for wideband channels, which is the main contribution of this paper.

\section{Review of OTFS Communication}\label{sec:OTFS}

OTFS converts a narrowband time-varying delay-spread wireless channel into a time-independent channel represented by a complex gain~\cite{Hadani2018Orthogonal, HadaniMar2017, Hadani2018otfs, HadaniIMS2017}. OFDM communication, on the other hand, converts a static (i.e., Doppler-free) multipath channel into a single tap channel, thus completely eliminating inter symbol interference (ISI). It has been shown that OTFS reduces to asymmetric OFDM (A-OFDM) in static multipath channels \cite{RavitejaJune2019}. 
\textcolor{black}{In narrowband time-varying delay-spread channels that arise in high mobility scenarios, OTFS receivers using turbo, message passing or MMSE equalizers outperform OFDM receivers using a sphere decoder or MMSE equalizer ~\cite{Hadani2018Orthogonal, HadaniMar2017, RavitejaJune2019}.}
We briefly describe the transmitter and receiver of an OTFS communication system and the propagation of the OTFS signal over narrowband time-varying delay-spread channels in the following subsections, primarily, to setup some notations in the paper.

\subsection{OTFS Transmitter}

\begin{figure}
	\setxysizeo
	\centering
	\includegraphics[scale=0.54]{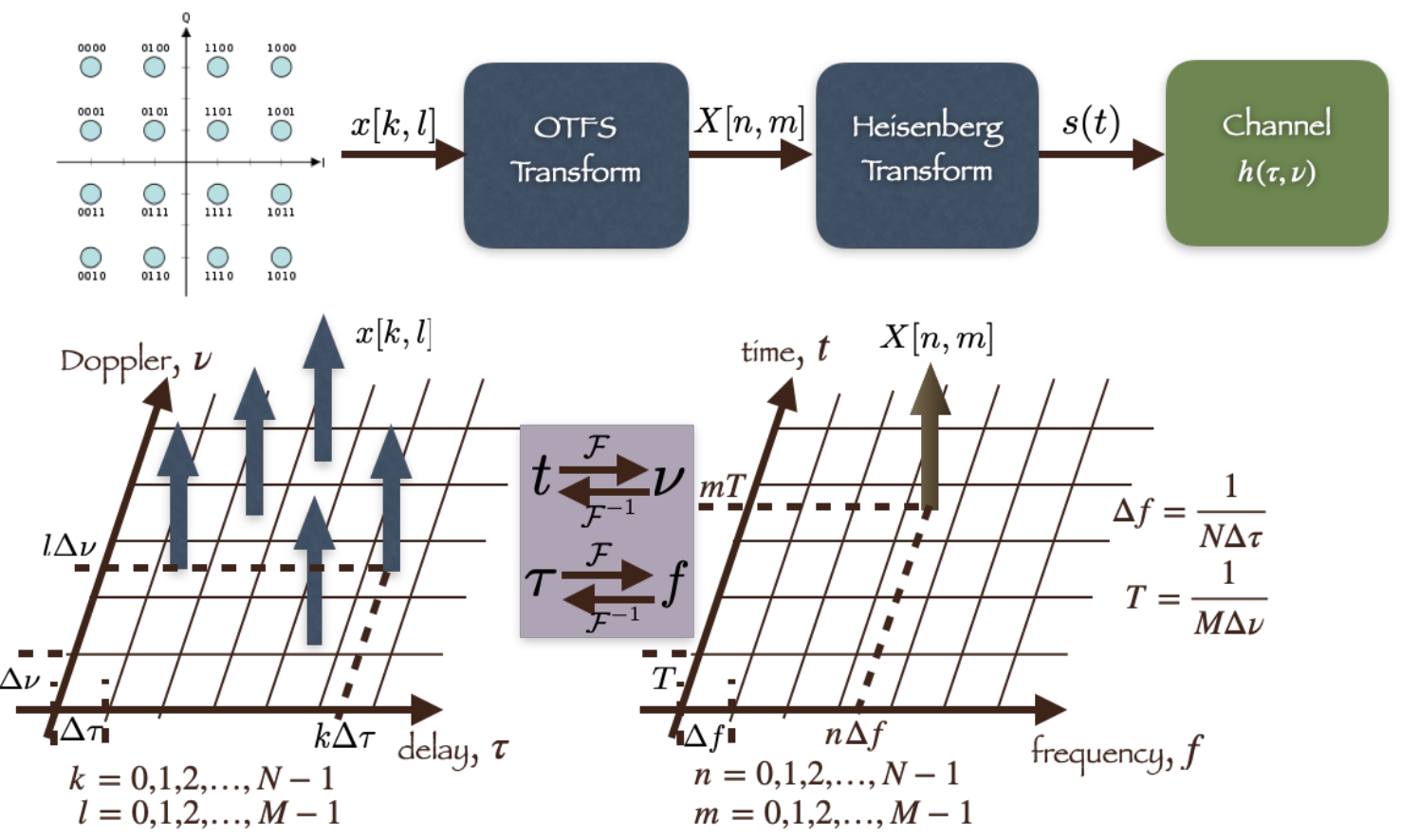}
	\caption{OTFS transmission scheme } \label{fig:otfsTx}
	\pullUp
	\vspace{-0.3cm}
\end{figure}

OTFS transmitter, shown in Fig. \ref{fig:otfsTx}, comprises the OTFS transform followed by the Heisenberg transform. The data (information bits) to be communicated, after bit-to-symbol mapping, are multiplexed onto a discrete 2D delay-Doppler domain grid of size $N \times M$. The OTFS transform maps the information symbols (e.g., QAM symbols), $x[k,l]$, in the discrete delay-Doppler space to the 2D sequence, $X[n,m]$, in the time-frequency domain by means of an inverse symplectic Fourier transform (ISFFT) as follows:
\begin{equation} \label{eqn:ISFFT}
X[n,m] = \frac{1}{NM} \sum_{l = 0}^{M-1} \sum_{k = 0}^{N-1} x[k,l] e^{j 2 \pi \left(  \frac{nl}{N} - \frac{mk}{M} \right)},
\end{equation}
where $m \in \lbrace{ 0,1,\ldots,M-1 \rbrace}, n \in \lbrace{0,1,\ldots,N-1 \rbrace}$. The $(N,M)$ periodized version of the input (respectively, output) 2D sequence, $x_p[k,l] \left( \text{resp. } X_p[n,m] \right)$, of the ISFFT reside on the lattice (reciprocal lattice), $\Lambda^\perp \triangleq \lbrace{ (k \Delta \tau, l\Delta \nu): k, l \in \mathbb{Z} \rbrace}$ $\left( \text{resp. }  \Lambda \triangleq \lbrace{ (mT, n \Delta f): m, n \in \mathbb{Z} \rbrace} \right)$, where $T$ and $\Delta f$ are the spacings on time and frequency axes, and
\begin{eqnarray}  \label{eqn:delFreqRes}
\Delta \tau = \frac{1}{N \Delta f}, \  
\Delta \nu = \frac{1}{MT},
\end{eqnarray}	
are the spacings on the delay and Doppler domain respectively. The Heisenberg transform converts the 2D time-frequency data, $X[n,m]$,  to a 1D continuous time-series, $s(t)$,  given by
\begin{equation}  \label{eqn:otfsModulator}
s(t) = \sum_{m=0}^{M-1} \sum_{n=0}^{N-1} X[n,m] e^{j2\pi n\Delta f(t-mT)} g_{\text{tx}}(t-mT),
\end{equation}	
where $g_{\text{tx}}(t)$ is the transmit pulse shaping function. We assume that the transmitted signal, $s(t)$, satisfies the narrowband assumption (A-1). 

The Heisenberg transform can be viewed as a map parameterized by the 2D time-frequency sequence, $X[n,m]$, and producing $s(t)$ when fed with $g_{\text{tx}}$, i.e., $s(t) = \Pi_X(g_{\text{tx}}(t))$: 
\begin{equation}  \label{eqn:otfsModulatorEquiv}
s(t) = \int_{\nu} \int_{\tau} X(\tau,\nu) e^{j2\pi\nu(t-\tau)} g_{\text{tx}}(t-\tau) d\tau d\nu,
\end{equation}	
where 
\begin{equation}\label{eqn:tfSignal}
X(\tau,\nu)  = \sum_{m=0}^{M-1} \sum_{n=0}^{N-1} X[n,m] \delta(\tau-mT,\nu-n\Delta f),
\end{equation}
with $\delta(\cdot)$ denoting the Dirac delta function. The above interpretation of the Heisenberg transform is helpful in relating the input and output of the OTFS system in the next subsection. 

\subsection{OTFS Signal Propagation}

\begin{figure}
	\setxysizeo
	\centering
	\includegraphics[scale=.54]{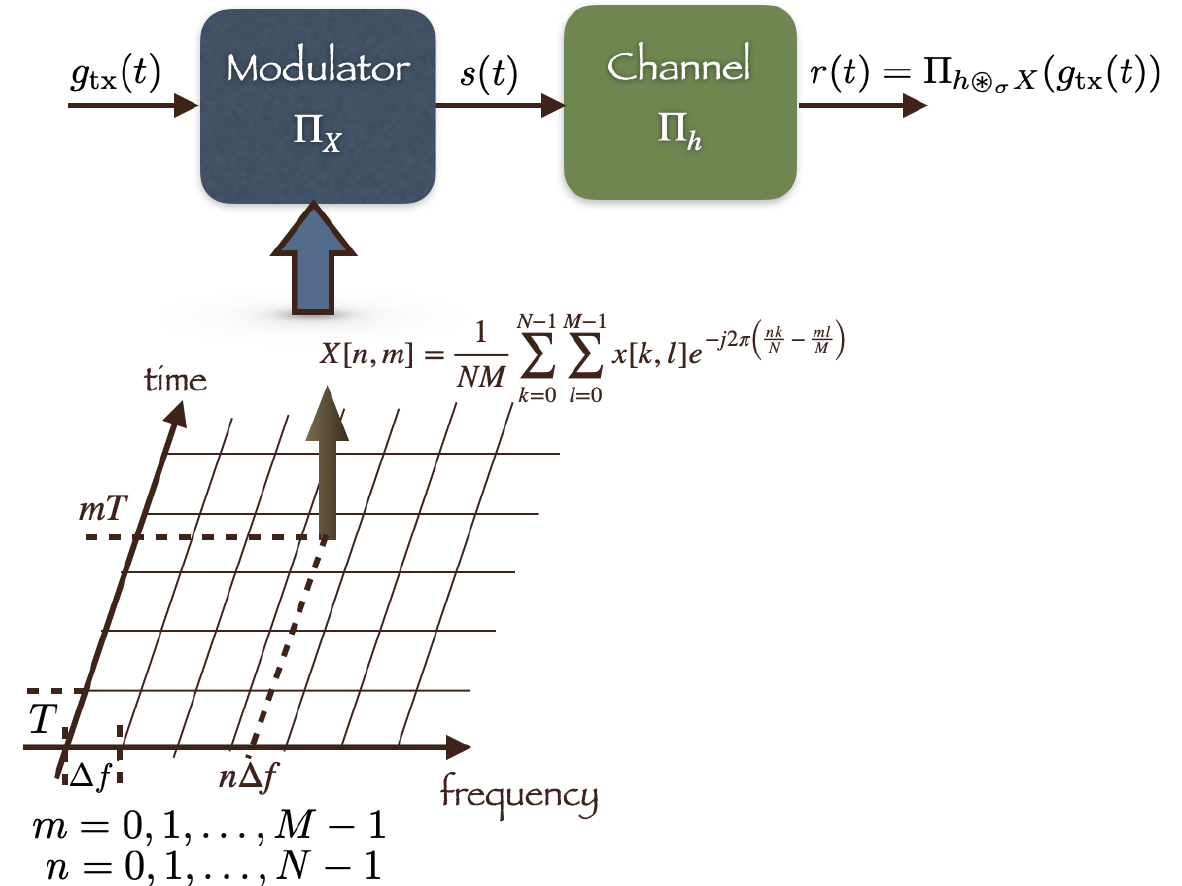}
	\caption{Cascade of OTFS modulator and propagation channel} \label{fig:otfsModChanRx}
	\pullUp
	\vspace{-0.3cm}	
\end{figure}

The signal, at the OTFS receiver, after propagating through a narrowband channel is given by \textcolor{black}{$r(t) = r_s(t) + w(t)$},
where $r_s(t)$ is the output of a narrowband channel as in \eqref{eqn:nbDelDoplChan} and $w(t)$ is the additive noise. As shown in \cite{Hadani2018Orthogonal}, the received signal, $r_s(t)$, can be expressed as
\begin{equation}  \label{eqn:otfsRxEquiv}
r_s(t) = \Pi_f(g_{\text{tx}}(t)) = \int_\nu \int_\tau f(\tau,\nu) e^{j2\pi\nu(t-\tau)} g_{\text{tx}}(t-\tau) d\tau d\nu,
\end{equation}	
where $f$ is the \emph{twisted convolution} of $h$ and $X$, denoted by $h  \circledast_\sigma X$,  defined as follows:
\begin{equation}  \label{eqn:twisted_conv}
f(\tau,\nu) = \int_{\nu'} \int_{\tau'} h(\tau',\nu') X(\tau-\tau',\nu-\nu') e^{j2\pi\nu'(\tau-\tau')} d\tau' d\nu',
\end{equation}	
which, due to \eqref{eqn:tfSignal}, can be written as a finite sum:
\begin{multline}  \label{eqn:twisted_conv_2}
f(\tau,\nu) = \sum_{n=0}^{N-1} \sum_{m=0}^{M-1} h(\tau-mT,\nu-n\Delta f) \\ \times X[n,m] e^{j2\pi(\nu-n\Delta f)mT}.
\end{multline}	

The received signal is, therefore, a result of passing the transmit pulse shaping function through an equivalent channel parameterized by the twisted convolution of the physical channel and the data dependent 2D time-frequency signal.  Fig.~\ref{fig:otfsModChanRx} depicts this interpretation. The signal received by the OTFS receiver, including the additive noise $w(t)$, is given by
\begin{equation}  \label{eqn:otfsRxSplusN}
r(t) = r_s(t) + w(t) = \Pi_{h  \circledast_\sigma X}(g_{\text{tx}}(t)) + w(t).
\end{equation}	

\subsection{OTFS Receiver}

The receiver performs OTFS demodulation followed by equalization and symbol decoding. OTFS demodulation is a two step process: discrete Wigner transform followed by symplectic Fourier transform (SFFT). The discrete Wigner transform (inverse of the discrete Heisenberg transform) is obtained by sampling the cross-ambiguity function between the received signal, $r(t)$, and a receive pulse shaping function, $g_{\text{rx}}(t)$. The demodulated time-frequency signal is given by
\begin{equation}  \label{eqn:otfsDWT}
\hat{Y}[n,m] = A_{g_{\text{rx}},r}(\tau, \nu)|_{\tau=mT, \nu=n\Delta f},
\end{equation}
where
\begin{multline}  \label{eqn:xAmbiguity}
A_{g_{\text{rx}},r}(\tau,\nu) \triangleq \int_t e^{-j2\pi\nu(t-\tau)} g^*_{\text{rx}}(t-\tau)  r(t) dt \\ = A_{g_{\text{rx}},r_s}(\tau,\nu) + A_{g_{\text{rx}},w}(\tau,\nu).
\end{multline}

It can be shown that
\begin{multline}  \label{eqn:otfsCrossAmbig}
A_{g_{\text{rx}},r_s}(\tau,\nu) = f(\tau,\nu) \circledast_\sigma A_{g_{\text{rx}},g_{\text{tx}}}(\tau, \nu) \\ = \sum_{n=0}^{N-1} \sum_{m=0}^{M-1} X[n,m] H_{n,m}(\tau, \nu),
\end{multline}
where
\begin{multline}  \label{eqn:Hnm}
H_{n,m}(\tau, \nu) \triangleq \int_{\nu''} \int_{\tau''}  h(\tau'',\nu'') e^{j2\pi \nu''mT} \\ \times A_{g_{\text{rx}},g_{\text{tx}}}(\tau-\tau''-mT,\nu-\nu''-n\Delta f) \\ \times e^{j2\pi (\nu''+n\Delta f) (\tau-\tau''-mT)} d\tau'' d\nu''. 
\end{multline}	

At this point, it is assumed that
\begin{enumerate}
	\item the channel response, $h(\tau,\nu)$, has a finite support bounded by $\left(\tau_{\max}, \nu_{\max}\right)$, and
	\item \emph{bi-orthogonality} of transmit and receive pulses holds in a robust manner \cite{Hadani2018Orthogonal}, i.e., the cross-ambiguity function vanishes in a neighborhood around the  \emph{non-zero} lattice points, $(mT, n\Delta f)$:  $A_{g_{\text{rx}},g_{\text{tx}}}(\tau,\nu) = 0$, for $\tau \in \left(mT - \tau_{\max}, mT + \tau_{\max} \right)$ and $\nu \in \left(n\Delta f - \nu_{\max}, n\Delta f + \nu_{\max} \right)$ except around the lattice point corresponding to $m=0, n=0$.
\end{enumerate}

Due to the above assumptions, upon sampling at $\tau = m_0T$ and $\nu = n_0 \Delta f$, we find $H_{n,m}[n_0, m_0]  = 0$ whenever $n \ne n_0$ or $m \ne m_0$, so that \eqref{eqn:otfsCrossAmbig} simplifies to
\begin{eqnarray}  \label{eqn:otfsCrossAmbig_}
A_{g_{\text{rx}},r_s}[n_0,m_0]  =  H_{n_0,m_0}[n_0, m_0]  X[n_0,m_0],
\end{eqnarray}	
where\footnote{Many papers on OTFS, either explicitly or implicitly, assume $A_{g_{\text{rx}},g_{\text{tx}}}(\tau,\nu) = 1$ in a neighborhood of $(\tau = 0, \nu=0)$ contained in the support of $h(\tau,\nu)$. \textcolor{black}{It is unclear if there are practical transmit and receive pulse shaping functions that satisfy this assumption, so we retain the term $A_{g_{\text{rx}},g_{\text{tx}}}(\tau,\nu)$ in the integrand in \eqref{eqn:Hnm0}}. } 
\begin{multline}  \label{eqn:Hnm0}
H_{n_0,m_0}[n_0, m_0] = \int_\nu \int_\tau  e^{-j2\pi \nu \tau}  h(\tau,\nu) \textcolor{black}{A_{g_{\text{rx}},g_{\text{tx}}}(\tau,\nu)} \\ \times e^{j2\pi (\nu m_0T - n_0\Delta f \tau)} d\tau d\nu.
\end{multline}	

Note that we choose $T \ge 2 \tau_{\max}$ and $\Delta f \ge 2 \nu_{\max}$ to avoid \textcolor{black}{ISI and ICI}, respectively, and hence ensure the validity of \eqref{eqn:otfsCrossAmbig_}. This means that, for a given $N$ and $M$ (and hence the number of symbols $NM$), the duration and bandwidth of the transmitted OTFS signal, $s(t)$, must be at least $2 M \tau_{\max}$ and $2 N \nu_{\max}$, respectively. The spectral efficiency of the OTFS scheme can, therefore, be at most $\frac{1}{4\tau_{\max} \nu_{\max}}$ symbols/s/Hz in a channel with a delay spread $\tau_{\max}$ and Doppler spread $\nu_{\max}$. These choices, according to \eqref{eqn:delFreqRes}, result in $\Delta \tau \le \frac{1}{2 N \nu_{\max}}$ and $\Delta \nu \le \frac{1}{2 M \tau_{\max}}$ in the delay-Doppler plane. Consider an OTFS transmit signal of bandwidth, $N \Delta f  = 2N \nu_{\max}$, which is the minimum required bandwidth to avoid ICI. As the narrowband assumption requires the signal bandwidth to satisfy $N \Delta f < \kappa f_c$, where $\kappa \ll 1$, we must choose $N < \kappa \frac{ f_c}{2 \nu_{\max}}$. Clearly, barring implementation aspects, there is no such upper limit on the choice of $M$.

The output of the discrete Wigner transform is, therefore, given by
\begin{equation}  \label{eqn:otfsDWTOutput}
\hat{Y}[n,m] = H_{n,m}[n, m]  X[n,m] + W[n,m],
\end{equation}
where $W[n,m]=A_{g_{\text{rx}},w}(\tau, \nu)|_{\tau=mT, \nu=n\Delta f}$ is the additive noise in the discrete time-frequency space. The OTFS demodulator output is obtained by taking SFFT of the discrete Wigner transform output:
\begin{eqnarray} 
\hat{y}[k,l] &=&  \sum_{m = 0}^{M-1} \sum_{n = 0}^{N-1} \hat{Y}[n,m] e^{-j 2 \pi \left(  \frac{nl}{N} - \frac{mk}{M} \right)},\\
&=& \sum_{m = 0}^{M-1} \sum_{n = 0}^{N-1} x[n,m] h_v\left(\frac{k-m}{MT}, \frac{l-n}{N \Delta f} \right) + w[k,l] \nonumber,  \label{eqn:otfsIO}\\
\end{eqnarray}
where $h_v(.,.)$ is obtained by sampling $\left( \nu =\frac{k-m}{MT}, \tau=\frac{l-n}{N \Delta f}\right)$ the function
\begin{eqnarray}  \label{eqn:hw}
h_v(\nu,\tau) &=& \sum_{n=0}^{N-1} \sum_{m=0}^{M-1} H_{n,m}[n,m] e^{j2 \pi \left(\tau n \Delta f - \nu m T\right)} \nonumber \\
&=&\iint  e^{-j2\pi \nu' \tau'} h(\tau',\nu') \textcolor{black}{A_{g_{\text{rx}},g_{\text{tx}}}(\tau',\nu')} \nonumber \\ 
& & \:\:\:\:\:\:\: \times v(\nu-\nu', \tau-\tau') d\tau' d\nu',
\end{eqnarray}
where $v(\nu, \tau) = \sum_{m=0}^{M-1} \sum_{n=0}^{N-1} e^{j 2\pi \left(\tau n \Delta f - \nu mT\right)}$ is a periodic function on the $\tau-\nu$ plane with periods $N\Delta \tau$ and $M \Delta \nu$ in delay and Doppler, respectively. 

\begin{figure}
	\setxysizeo
	\centering
	\includegraphics[scale=.72]{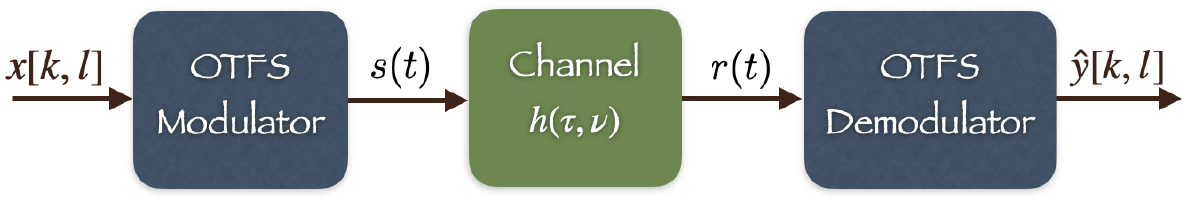}
	\caption{OTFS Block Diagram} \label{fig:otfsHighLevelBlockDiag}
	\pullUp
	\vspace{-0.3cm}	
\end{figure}

Equation \eqref{eqn:otfsIO} depicts the input-output relation in an OTFS system that can be written in the following vectorized form:
\begin{equation}  \label{eqn:otfsIO_vector}
\mathbf{y} = \mathbf{H} \mathbf{x} + \mathbf{w},
\end{equation}
where $\mathbf{y} \in \mathbb{C}^{NM \times 1}$ is the output of the OTFS demodulator whose $(k + Nl)$-th entry is $y[k,l]$,  $\mathbf{H} \in \mathbb{C}^{NM \times NM}$ is the effective channel matrix, $\mathbf{x} \in \mathbb{C}^{NM \times 1}$ is the symbol vector whose $(k + Nl)$-th entry is $x[k,l]$ and $\mathbf{w} \in \mathbb{C}^{NM \times 1}$ is the additive noise at the OTFS demodulator output. Fig. \ref{fig:otfsHighLevelBlockDiag} summarizes various stages at a block level from the input to output of an OTFS system.

\textcolor{black}{Equalization and symbol decoding is performed after OTFS demodulation to recover the transmitted information bits.} In OTFS modulation, developed for the narrowband channel model in \eqref{eqn:nbDelDoplChan}, the information symbols were mounted on the delay-Doppler grids. Fourier and inverse Fourier transforms were made use of to move between $\tau$ and $f$ domains, and between $t$ and $\nu$ domains, respectively. In the new ODSS modulation, to be developed for the wideband channel model in \eqref{eqn:wbDelScaleChan}, we shall make use of the Mellin and inverse Mellin transforms to move between the scale ($\alpha$) domain and Mellin ($\beta$) domain, respectively. We next present the Mellin transform and its discrete counterpart before we develop the ODSS modulation scheme for wideband delay-scale channels.

\section{Mellin Transform and its Properties} \label{sec:MellinX}

The Mellin transform was developed in \cite{BertrandBook} as a solution to the problem of finding the transform that enjoys a \emph{scale-invariance} property, i.e., the Mellin transform of the signal $\sqrt{a} x(a \alpha ), a>0, \alpha >0,$ is same as that of the original signal, $x(\alpha)$, except for a phase shift. The Mellin transform of a signal $x(\alpha), \alpha>0$, is defined by
\begin{equation}  \label{eqn:mellin}
\mathcal{M}_x(\beta) \triangleq \int_{0}^{\infty} \frac{1}{\sqrt{\alpha}} x(\alpha) e^{j2\pi \beta \log(\alpha)} d\alpha,
\end{equation}
where $\beta \in \mathbb{R}$ is the Mellin variable. The Mellin transform exists for signals, $x(\alpha)$, in the \textcolor{black}{Hilbert space $\mathcal{L}^2\left(\mathbb{R}^+, \frac{d \alpha}{\sqrt{\alpha}} \right)$ which is the set of square integrable functions ($\mathcal{L}^2$-functions) over $\mathbb{R}^+$, attached with a measure $\frac{d \alpha}{\sqrt{\alpha}}$ instead of $d\alpha$.} We may interpret Mellin transform as the Fourier transform of the function $\sqrt{e^{t}} x(e^{t}), t \in \mathbb{R}$ \cite{Antonio}. 

We state a few important properties \textcolor{black}{that make Mellin transform  attractive for working with time-scaling effect of Doppler in wideband channels~\cite{BertrandBook, Papandreou}.}
\begin{enumerate}
 \item The scale-invariance property of the Mellin transform follows from the definition in \eqref{eqn:mellin}: the Mellin transform of the scaled version $\sqrt{a} x(a \alpha ), a>0,$ of $x(\alpha)$ is given by $a^{-j 2 \pi \beta} \mathcal{M}_x(\beta) $ which is the same as the Mellin transform of the original signal except for a phase shift. Note that this scale-invariance property parallels the shift-invariance property of the Fourier transform: the Fourier transform of $x(t - \tau)$ is $e^{-j2\pi f \tau} X(f)$ where $X(f)$ is the Fourier transform of $x(t)$  \cite{Oppenheim}.
 
 \item The Mellin transform of the dilation-invariant product of two functions $x_1(\alpha)$ and $x_2(\alpha)$, defined by $(x_1 \circ x_2)(\alpha) = \sqrt{\alpha} x_1(\alpha) x_2(\alpha)$, is given by the linear convolution of the Mellin transforms of the two functions: $\mathcal{M}_{x_1 \circ x_2}(\beta) = \mathcal{M}_{x_1}(\beta) \circledast \mathcal{M}_{x_2}(\beta)$.
 
 \item The geometrical Dirac comb, $\Delta_A(\alpha) = \sum_{n=-\infty}^{\infty} A^{n/2} \delta(\alpha - A^n)$, in the scale-space, $\mathbb{R}^+$,  transforms to an arithmetical comb in the Mellin space, $\mathcal{M}_{\Delta_A}(\beta) =\frac{1}{\ln A} \sum_{n=-\infty}^{\infty} \delta(\beta - \frac{n}{\ln A})$.
 
 \item Parseval's theorem: $\int_{-\infty}^{\infty} \mathcal{M}_{x_1}(\beta) \mathcal{M}_{x_2}^*(\beta) d\beta = \int_{0}^{\infty} x_1(\alpha) x_2^*(\alpha) d\alpha$.
 
 \item The Mellin transform of the multiplicative convolution of two functions $x_1(\alpha)$ and $x_2(\alpha)$, defined by $(x_1 \vee x_2)(\alpha) = \int_{0}^{\infty}  \sqrt{\alpha} x_1(\alpha') x_2(\frac{\alpha}{\alpha'}) \frac{d\alpha'}{\alpha'}$, is given by the product of the Mellin transforms of the two functions: $\mathcal{M}_{x_1 \vee x_2}(\beta) = \mathcal{M}_{x_1}(\beta)  \mathcal{M}_{x_2}(\beta)$.
 
 \item The inverse Mellin transform is given by
 \begin{equation}  \label{eqn:iMellin}
 x(\alpha) \triangleq \frac{1}{\sqrt{\alpha}} \int_{\textcolor{black}{-\infty}}^{\infty} \mathcal{M}_x(\beta) e^{-j2\pi \beta \log(\alpha)} d\beta, \alpha>0,
 \end{equation}
 which follows immediately upon noting the relation between the Mellin and Fourier transforms.
 
\end{enumerate}

We now consider the effect of discretization on Mellin transform relations \cite{BertrandBook, Bertrand1990}. First, consider geometric sampling in the scale  domain ($\alpha$-domain) with ratio $q$. The Mellin transform of the sampled version of $x(\alpha)$, i.e., $x_s(\alpha) \triangleq \left(x \circ \Delta_q\right)(\alpha) = \sum_{n=-\infty}^{\infty} q^{n/2} x(q^n) \delta(\alpha - q^n)$, is given by
\begin{eqnarray}  \label{eqn:periodizedMellin}
\mathcal{M}_{x_s}(\beta) &=& \mathcal{M}_{x \circ \Delta_q}(\beta) \nonumber\\
&\overset{(a)}{=}& \mathcal{M}_{x}(\beta) \circledast \mathcal{M}_{\Delta_q}(\beta) \nonumber\\
&\overset{(b)}{=}& \frac{1}{\ln q} \sum_{n=-\infty}^{\infty} \mathcal{M}_{x}\left(\beta - \frac{n}{\ln q}\right)  \\
&\triangleq& \mathcal{M}_{x}^P(\beta),
\end{eqnarray}
where the step (a) follows from the fact that Mellin transform of a dilation-invariant product in the scale domain corresponds to the convolution of Mellin transforms in the Mellin domain, and step (b) follows upon an evaluation of the convolution in step (a). Therefore, geometric sampling in the scale domain leads to periodization in the Mellin domain. Aliasing due to scale domain geometric sampling is avoided if: 
\begin{enumerate}
	\item the Mellin spectrum is $\beta$-limited, i.e.,  $\mathcal{M}_{x}(\beta)$ is nonzero only in a finite interval $\left[\beta_1, \beta_2\right]$, and 
	
	\item the geometric sampling ratio, $q$, satisfies 
	\begin{equation}\label{eqn:qLimit}
		\frac{1}{\ln q} \ge \beta_2 - \beta_1.
	\end{equation}
\end{enumerate}

Next, consider  sampling the Mellin domain function, $\mathcal{M}_{x}(\beta)$. Sampling in the Mellin space results in
\begin{eqnarray}  \label{eqn:samplingMellin}
M_s(\beta) &\triangleq& \frac{1}{\ln Q} \sum_{n=-\infty}^{\infty} \mathcal{M}_{x}\left(\frac{n}{\ln Q}\right) \delta\left(\beta - \frac{n}{\ln Q}\right) \nonumber\\
\\
&\overset{(c)}{=}& \mathcal{M}_{x}(\beta) \mathcal{M}_{\Delta_Q}(\beta)  \nonumber\\
&\overset{(d)}{=}& \mathcal{M}_{x \vee \Delta_Q}(\beta),
\end{eqnarray}
where equality (c) follows from the formula for the Mellin transform of a geometric impulse train in the scale domain, and (d) follows from the fact that multiplicative convolution in scale domain corresponds to the product of the Mellin transforms. We see that the sampled version of the Mellin transform of $x(\alpha)$, i.e., $M_s(\beta)$, is the inverse Mellin transform of the \emph{dilatocycled}  version of $x(\alpha)$ given by 
\begin{equation}
x_d(\alpha) \triangleq \left(x \vee \Delta_Q\right)(\alpha) = \sum_{n=-\infty}^{\infty} Q^{n/2} x(Q^n \alpha).
\end{equation}
Thus, sampling in the Mellin domain leads to dilatocycling in the scale domain. Aliasing due to Mellin domain sampling is avoided if:
\begin{enumerate}
	\item the signal in the scale domain has a finite support, $\left[\alpha_1, \alpha_2\right]$,  and 
	
	\item the dilatocycling ratio, $Q$, satisfies: $Q \ge \frac{\alpha_2}{\alpha_1}$.
\end{enumerate}
In the absence of aliasing, $x_d(\alpha)$ equals $x(\alpha)$ for $\alpha \in \left[\alpha_1, \alpha_2\right]$.

Finally, the discrete Mellin transformation is obtained by geometric sampling of the finitely supported and dilatocycled signal $x_d(\alpha), \alpha \in \left[\alpha_1, \alpha_2\right]$, in the scale domain. The sampled version of $x_d(\alpha)$ is given by
\begin{equation}
x_{ds}(\alpha) \triangleq \left(x_d \circ \Delta_q\right)(\alpha) = \sum_{n=-\infty}^{\infty} q^{n/2} x_d(q^n) \delta(\alpha - q^n).
\end{equation}

 It is clear from the discussions above that the Mellin transform of $x_{ds}(\alpha)$ is the periodized version of $M_s(\beta)$:
\begin{equation}  \label{eqn:sampledPeriodizedMellin}
\mathcal{M}_{x_{ds}}(\beta)=M_s^P(\beta) \triangleq \frac{1}{\ln q} \sum_{n=-\infty}^{\infty} M_s\left(\beta - \frac{n}{\ln q}\right),
\end{equation}
where we require $\frac{1}{\ln q} \ge \beta_2 - \beta_1$ to avoid aliasing. Substituting $M_s(\beta)$ from \eqref{eqn:samplingMellin} in \eqref{eqn:sampledPeriodizedMellin} and restricting $Q = q^N$, where $N$ is a positive integer, we get:
\begin{multline}  \label{eqn:sampledPeriodizedMellinSub}
M_s^P(\beta) = \frac{1}{\ln q \ln Q } \sum_{m=-\infty}^{\infty} \sum_{n=-\infty}^{\infty} \mathcal{M}_{x}\left(\frac{m}{N \ln q}\right) \\ \times \delta\left(\beta - \frac{nN+m}{\ln q}\right).
\end{multline}
Changing $m \rightarrow k = m + nN$ and using the definition of periodized version, we get
\begin{equation}  \label{eqn:sampledPeriodizedMellinSub_}
M_s^P(\beta) = \frac{1}{\ln Q } \sum_{k=-\infty}^{\infty} \mathcal{M}_{x}^P\left(\frac{k}{N \ln q}\right) \delta\left(\beta - \frac{k}{\ln q}\right).
\end{equation}

It is now straightforward to show that the discrete Mellin transform relationship is given by
\begin{equation}  \label{eqn:DMT}
\mathcal{M}_{x}^P\left(\frac{k}{\ln Q}\right) = \sum_{n=J}^{J+N-1}  q^{n/2} x_d(q^n) e^{j2 \pi nk/N},
\end{equation}
where $J$ is the integer part of $\ln \alpha_1/\ln q$.  The transform length $N = \frac{\ln Q}{\ln q}$ must satisfy the condition
\begin{equation}  \label{eqn:nLimit}
N \ge \left( \beta_2 - \beta_1 \right) \ln\left( \frac{\alpha_2}{\alpha_1} \right),
\end{equation}
to avoid aliasing and allow reconstruction of the scale and Mellin domain functions from their samples.

Similarly, the discrete inverse Mellin transform is given by
\begin{equation}  \label{eqn:iDMT}
x_d(q^n) =  \frac{q^{-n/2}}{N} \sum_{k=K_i}^{K_i+N-1}   \mathcal{M}_{x}^P\left(\frac{k}{\ln Q}\right) e^{-j2 \pi kn/N},
\end{equation}
where $K_i$ is the integer part of $\beta_1 \ln Q$. 

\section{ODSS Communication For Wideband Channels}\label{sec:ODSS}

We now turn to developing the ODSS modulation. The goal of ODSS modulation is to convert a  wideband, time-varying, delay-scale spread channel into a time-independent channel represented by a complex gain. To this end, we introduce the 2D \emph{ODSS transform} (and its inverse) which is a combination of discrete Fourier transform on one axis (the delay axis) and inverse Mellin transform on the other (the scale axis.) The development of ODSS parallels the development of OTFS in Section \ref{sec:OTFS}. 
\textcolor{black}{In the process, we appropriately modify the two key properties -- twisted convolution property and  robust biorthogonality} -- that were used in the development of OTFS. We develop the transmitter and receiver of an ODSS communication system and the propagation of the signal over wideband time-varying channels in the following subsections, which is the main contribution of this paper. \textcolor{black}{While we develop ODSS in a manner similar to the development of OTFS, we note that the two modulation schemes are distinct and do not generalize or reduce to each other.}

\subsection{ODSS Transmitter}

The information bits, after bit-to-symbol mapping, are multiplexed onto the discrete 2D Mellin-Fourier domain of size, $M_{\text{tot}}=\sum_{n=0}^{N-1} M(n)$, where $M(n) = \lfloor q^n \rfloor$. The ODSS transform maps the data symbols (e.g., QAM symbols), $\lbrace x[k,l]: k = 0,1,\ldots,N-1,  l=0,1,\ldots,M(k)\rbrace$, in the discrete Mellin-Fourier space to the 2D sequence, $X[n,m]$, in the scale-delay domain by taking an inverse discrete Mellin transform along the scale axis (see \eqref{eqn:iDMT}) and a discrete Fourier transform along the delay axis, as follows:
\textcolor{black}{\begin{equation} \label{eqn:ODSStransform}
X[n,m] =  \frac{q^{-n/2}}{N} \sum_{k=0}^{N-1}  \frac{\sum_{l = 0}^{M(k)-1}  x[k,l] e^{j 2 \pi \left(  \frac{ml}{M(k)} - \frac{nk}{N} \right)}}{M(k)} ,
\end{equation}}
where $m \in \lbrace{ 0,1,\ldots,M(n)-1 \rbrace}, n \in \lbrace{0,1,\ldots,N-1 \rbrace}$. 
The  periodized version of the input (respectively, output) 2D sequence, $x_p[k,l] \left( \text{resp. } X_p[n,m] \right)$, reside on the lattice (reciprocal lattice), $\Lambda^\perp = \lbrace{ (k \Delta \beta, l\Delta f): k, l \in \mathbb{Z} \rbrace}$ $\left( \text{resp. }  \Lambda = \lbrace{ (m \Delta \tau, q^n): m, n \in \mathbb{Z} \rbrace} \right)$, where $\Delta \beta = \frac{1}{N\ln q}$, $\Delta f$, $\Delta \tau$ are the spacings on the Mellin, Fourier and delay axes, respectively; $\Delta \tau = \frac{1}{ W}$, $W \triangleq  M \Delta f $, and  $q$ is the geometric sampling ratio on the scale axis. The sampling ratio, $q$, and discrete Mellin transform length, $N$, are chosen to satisfy the conditions in \eqref{eqn:qLimit} and \eqref{eqn:nLimit}. \textcolor{black}{We may express \eqref{eqn:ODSStransform} in the vectorized form:} 
\textcolor{black}{\begin{equation} \label{eqn:ODSStransformVectorForm}
		\mathbf{X} = \mathcal{T}_{\text{iMF}} \mathbf{x},
\end{equation}}
\textcolor{black}{where $\mathbf{x} \in \mathbb{C}^{M_{\text{tot}} \times 1}$ is the symbol vector obtained by stacking $x[k,l]$ into a vector, $\mathbf{X} \in \mathbb{C}^{M_{\text{tot}} \times 1}$ is the vector obtained by stacking $X[n,m]$, and $\mathcal{T}_{\text{iMF}} \in \mathbb{C}^{M_{\text{tot}} \times M_{\text{tot}}}$ is the matrix representing the 2D ODSS transform in \eqref{eqn:ODSStransform}.}

\begin{figure}
	\setxysizeo
	\centering
	\includegraphics[scale=.42]{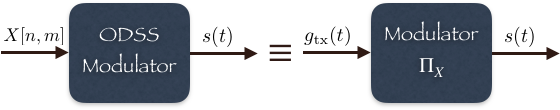}
	\caption{ODSS Modulator Representation} \label{fig:odss_Modulator}
	\pullUp
	\vspace{-0.3cm}	
\end{figure}

The ODSS modulator converts the 2D time-frequency data, $X[n,m]$,  to a 1D continuous time-series, $s(t)$,  given by
\textcolor{black}{\begin{equation}  \label{eqn:odssModulator}
s(t) = \sum_{n=0}^{N-1} \sum_{m=0}^{M(n)-1}  X[n,m] q^{n/2} g_{\text{tx}} \left(q^n \left(t-\frac{m}{q^nW}\right)\right),
\end{equation}
where $g_{\text{tx}}(t)$ is the transmit pulse shaping function of duration $T = 1/W$.} The ODSS modulation can be viewed as a map parameterized by the 2D Mellin-Fourier sequence, $X[n,m]$, and producing $s(t)$ when fed with $g_{\text{tx}}$, i.e., $s(t) = \Pi_X(g_{\text{tx}}(t))$: 
\begin{equation}  \label{eqn:odssModulatorEquiv}
s(t) =  \iint  X(\tau,\alpha) \sqrt{\alpha} g_{\text{tx}}(\alpha(t-\tau)) d\tau d\alpha,
\end{equation}	
where 
\textcolor{black}{\begin{equation}\label{eqn:sdSignal}
X(\tau,\alpha)  = \sum_{n=0}^{N-1} \sum_{m=0}^{M(n)-1}  X[n,m] \delta(\tau-\textcolor{black}{\frac{m}{q^nW}},\alpha-q^n).
\end{equation}}
The above interpretation of the ODSS transform, depicted in Fig. \ref{fig:odss_Modulator}, is helpful in relating the input and output of an ODSS communication system in the next subsection. 

\subsection{ODSS Signal Propagation}

\begin{figure}
	\setxysizeo
	\centering
	\includegraphics[scale=.42]{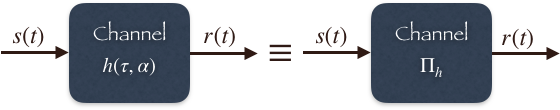}
	\caption{Wideband Channel Representation} \label{fig:DSchannel}
	\pullUp
	\vspace{-0.3cm}	
\end{figure}

The signal, at the ODSS receiver, after propagating through a wideband delay-scale channel is given by \textcolor{black}{$r(t) = r_s(t) + w(t)$},
where $r_s(t)$ is as in \eqref{eqn:wbDelScaleChan}, and $w(t)$ is the additive noise.

We may, equivalently, view the propagation channel as performing the map {$\Pi_h(s): s(t) \rightarrow r_s(t)$} as shown in Fig.~\ref{fig:DSchannel}. Next, we introduce the notion of $\omega$-convolution to describe the equivalent of the cascade of the ODSS modulator and the propagation channel.

\begin{figure}[t]
	\setxysizeo
	\centering
	\includegraphics[scale=.42]{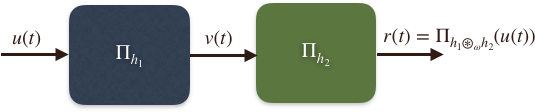}
	\caption{$\omega$-convolution} \label{fig:cascade_DS_channel}
	\pullUp
	\vspace{-0.1cm}	
\end{figure}

The cascade of two delay-scale channels, as shown in Fig.~\ref{fig:cascade_DS_channel}, is equivalent to a single channel, i.e., $\Pi_{h_2}(\Pi_{h_1}(s)) = \Pi_{h}(s)$, where $h(\tau,\alpha) = h_2(\tau,\alpha) \circledast_\omega  h_1(\tau,\alpha)$ and the symbol $\circledast_\omega$ denotes the $\omega$-convolution defined by
\begin{equation}  \label{eqn:omega_conv}
h(\tau,\alpha) = \iint h_2(\tau',\alpha') h_1\left(\alpha'(\tau-\tau'),\frac{\alpha}{\alpha'}\right) d\tau' d\alpha'.
\end{equation}	

The derivation of the above is provided in \textcolor{black}{\cite[Sec. \ref{supp:proof_eqn_omega_conv}]{p2021orthogonal}.} 


In light of the above result, we may write the signal after propagation through the channel, $r_s(t)$, as 
\begin{equation}  \label{eqn:odssRx_Equiv}
r_s(t) = \Pi_{h \circledast_\omega X}(g_{\text{tx}}) = \iint f(\tau,\alpha)  \sqrt{\alpha}  g_{\text{tx}}(\alpha(t-\tau)) d\tau d\alpha \nonumber
\end{equation}	
where $f(\tau,\alpha)$ is given by (see \textcolor{black}{\cite[Sec. \ref{supp:proof_odssTransform}]{p2021orthogonal}}):
\begin{equation}  \label{eqn:odssTransform}
f(\tau,\alpha) =  \sum_n \sum_m X[n,m]  h\left(\tau-\textcolor{black}{\frac{m}{\alpha W}}, \frac{\alpha}{q^n} \right) \textcolor{black}{q^{-n}}. \nonumber \\ 
\end{equation}

\begin{figure}
	\setxysizeo
	\centering
	\includegraphics[scale=.36]{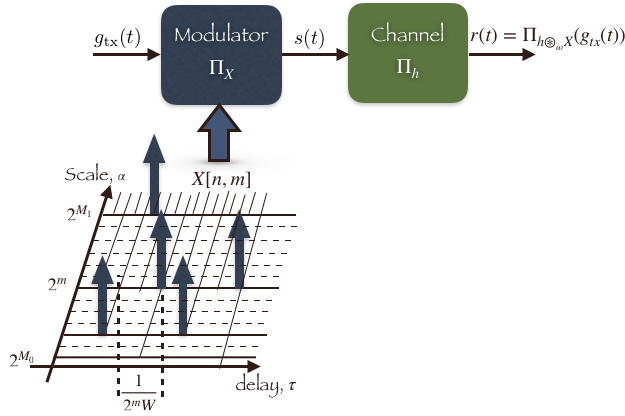}
	\caption{Received ODSS signal} \label{fig:odssModulator_channel}
	\pullUp
	\vspace{-0.3cm}	
\end{figure}

The received signal is, therefore, a result of passing the transmit pulse shaping function through an equivalent channel parameterized by the $\omega$-convolution of the physical channel and the data dependent 2D delay-scale signal.  Fig. \ref{fig:odssModulator_channel} depicts this interpretation. The signal received by the ODSS receiver, including the additive noise $w(t)$, is given by
\begin{equation}  \label{eqn:odssRxSplusN}
r(t) = r_s(t) + w(t) = \Pi_{h  \circledast_\omega X}(g_{\text{tx}}(t)) + w(t).
\end{equation}	

\subsection{ODSS Receiver}\label{subsec:odssReceiver}
The ODSS receiver performs ODSS demodulation followed by equalization and symbol decoding. ODSS demodulation is a two step process: extracting the transmitted scale-delay signal followed by an inverse ODSS transform. We describe the two steps in the following two subsections. 

\subsubsection{Scale-delay signal extraction} The scale-delay signal is extracted by sampling the cross-ambiguity function between the received signal and the pulse shaping function at the receiver side. The demodulated scale-delay signal is given by
\textcolor{black}{\begin{equation}  \label{eqn:odssDWT}
\hat{Y}[n,m] = A_{g_{\text{rx}},r}(\tau, \alpha)|_{\tau=\frac{m}{q^nW}, \alpha=q^n},
\end{equation}}
where
\begin{eqnarray}  \label{eqn:Xambiguity}
A_{g_{\text{rx}},r}(\tau,\alpha) &\triangleq& \int g^*_{\text{rx}}\left(\alpha(t-\tau)\right)  \sqrt{\alpha} r(t) dt \nonumber\\ 
&=& A_{g_{\text{rx}},r_s}(\tau,\alpha) + A_{g_{\text{rx}},w}(\tau,\alpha).
\end{eqnarray}
It can be shown (see \textcolor{black}{\cite[Sec. \ref{supp:proof_odssIO}]{p2021orthogonal}}) that
\begin{equation}  \label{eqn:odssIo}
A_{g_{\text{rx}},r_s}(\tau,\alpha)  =  \sum_n \sum_m X[n,m] H_{n,m}(\tau, \alpha),
\end{equation}	
where
\begin{multline}  
H_{n,m}(\tau, \alpha) = \iint h(\tau'',\alpha'')  \\ 
\times A_{g_{\text{rx}},g_{\text{tx}}}\left(\alpha'' q^n \left( \tau-\textcolor{black}{\frac{m}{\alpha'' q^n W}}-\tau''\right), \frac{\alpha}{\alpha''q^n}\right) d\tau'' d\alpha''. \nonumber 
\end{multline}

We assume that 
\begin{enumerate}
	\item the channel response has a finite support, i.e., $h(\tau,\alpha)$ is non-zero only for  $-\tau_{\max} \le \tau \le \tau_{\max}$ and $\frac{1}{\alpha_{\max}} \le \alpha \le  \alpha_{\max}$, where $\alpha_{\max} \ge 1$, and
	\item \emph{robust bi-orthogonality} holds between the transmit and receive pulses in the following manner. The cross-ambiguity function vanishes in the neighborhood of all lattice points $(\frac{m}{q^n W}, q^{n})$ \emph{except} $(0,1)$ corresponding to $m=0$ and $n=0$. That is,  $A_{g_{\text{rx}},g_{\text{tx}}}(\tau, \alpha)=0$ for $\tau \in (\frac{m}{q^n W}  - \tau_{\max}, \frac{m}{q^n W} + \tau_{\max})$ and $\alpha \in (q^{n}/\alpha_{\max}, q^{n} \alpha_{\max})$ except when $m=0$ and $n=0$.\footnote{Bi-orthogonality cannot be satisfied exactly; our choice of waveforms for ODSS implementation is discussed in Section \ref{sec:pulseShapeFn}}
\end{enumerate}

Then, on sampling at $\tau = \frac{m_0}{q^{n_0} W}$ and $\alpha =q^{n_0}$, we find that $H_{n,m}[n_0, m_0]  = 0$ whenever $n \ne n_0$ or $m \ne m_0$, and
\begin{multline}   \label{eqn:Hnm0_}
H_{n_0,m_0}[n_0, m_0] = \iint  h(\tau',\alpha')  \\ 
\times A_{g_{\text{rx}},g_{\text{tx}}}\left(q^{n_0} \left(\frac{m_0}{q^{n_0} W}\left(\alpha'-1\right) - \alpha' \tau'\right),\frac{1}{\alpha'}\right) d\tau' d\alpha'.
\end{multline}
so that the noise free part of the extracted scale-delay signal is given by
\begin{eqnarray}  \label{eqn:odssIo_}
A_{g_{\text{rx}},r_s}[n_0,m_0]  &=&  H_{n_0,m_0}[n_0, m_0]  X[n_0,m_0].
\end{eqnarray}

Consider, for example, a channel without delay and Doppler spread: $h(\tau,\alpha) = h_0 \delta(\tau,\alpha-1)$. In this case, we find: $H_{n,m}(\tau, \alpha) = A_{g_{\text{rx}},g_{\text{tx}}}\left(\tau-\frac{m}{q^n W},\frac{\alpha}{q^n}\right)$. Upon sampling at $\tau = \frac{m_0}{q^{n_0} W}$ and $\alpha = q^{n_0}$, due to robust bi-orthogonality, $H_{n,m}[n_0, m_0]  = 0$ whenever $n \ne n_0$ or $m \ne m_0$, and 
\begin{equation}  \label{eqn:Hnm0_AWGN}
H_{n_0,m_0}[n_0, m_0] =  h_0 A_{g_{\text{rx}},g_{\text{tx}}}\left(0 ,1\right) = h_0\nonumber,
\end{equation}
so that, in this special case, the noise free part of the extracted scale-delay signal is given by
\begin{equation}
A_{g_{\text{rx}},r_s}[n_0,m_0]  =  h_0 X[n_0,m_0].
\end{equation}
Therefore, for an ideal channel without delay and Doppler spread, the ODSS scheme produces a constant gain for all signal components in the extracted delay-scale domain. 

In general, we find from \eqref{eqn:odssIo_} that the ODSS scheme leads to an ISI free, time-independent, \textcolor{black}{scalar complex channel gain for each \emph{delay-scale} domain output at the receiver.} The extracted delay-scale signal at the ODSS receiver is, therefore, given by
\begin{equation}  \label{eqn:odssDWTOutput}
\hat{Y}[n,m] = H_{n,m}[n, m]  X[n,m] + W[n,m],
\end{equation}
where \textcolor{black}{$W[n,m]=A_{g_{\text{rx}},w}(\tau, \alpha)|_{\tau=\frac{m}{q^nW}, \alpha=q^n}$} is the additive noise in the discrete delay-scale space.

\textcolor{black}{To avoid ICI}, and hence obtain \eqref{eqn:odssIo_}, we need to 
\begin{enumerate}
	\item choose $q$ such that:
	\begin{equation}\label{eqn:c1}
		\frac{q^{n'}}{\alpha''q^n} \notin \left(\alpha_{\max}^{-1} , \alpha_{\max}\right),
	\end{equation}
 	$\forall \alpha'' \in \left(\alpha_{\max}^{-1}, \alpha_{\max}\right)$, whenever $n' \ne n$, and
	\item choose $q$ and $W$ such that:
	\begin{equation}\label{eqn:c2}
		\alpha'' q^n \left( \frac{m'}{W}-\frac{m}{\alpha'' q^n W}-\tau''\right) \notin \left(-\tau_{\max},\tau_{\max}\right),
	\end{equation}
	$\forall \tau'' \in \left(-\tau_{\max},\tau_{\max}\right)$ and $\alpha'' \in \left(\alpha_{\max}^{-1}, \alpha_{\max}\right)$, whenever $m' \ne m$.
\end{enumerate}

\textcolor{black}{We first choose the geometric sampling ratio, $q$, to meet the condition in \eqref{eqn:c1}. If $n'>n$, we want $q^{n'-n} \ge \alpha_{\max} \alpha''$, which is satisfied if: $\forall n' > n, q^{n'-n} \ge \alpha_{\max}^2$, i.e., if $q \ge \alpha_{\max}^2$. Similarly, if $n' < n$, we require $q^{n'-n} \le \alpha''\alpha_{\max}^{-1}$ which is met if: $\forall n' < n, q^{n'-n} \le \alpha_{\max}^{-2}$, i.e., if $q \ge \alpha_{\max}^2$. Therefore, we may choose
	\begin{equation}
		q = \alpha_{\max}^2. \label{eqn:qChoice}
	\end{equation}
Clearly, since $\alpha_{\max} \ge 1$, we have $q \ge 1$. Aside, we also note that the choice of $q$ in \eqref{eqn:qChoice} together with the robust bi-orthogonality property renders $A_{g_{\text{rx}},g_{\text{tx}}}(\tau, \alpha)=0$, $\alpha \notin \left(\alpha_{\max}^{-1} , \alpha_{\max}\right)$.}

\textcolor{black}{Next, with $q$ as in \eqref{eqn:qChoice}, we choose $W$ to satisfy the condition in \eqref{eqn:c2}. The condition in \eqref{eqn:c2} is equivalent to 
\begin{equation} \label{eqn:cond2Eq}
\underset{ \left(\tau'', \alpha'' \right) \in S}{\inf} \:  \left| \alpha'' \textcolor{black}{\alpha_{\max}^{2n}} m' - m - \alpha'' \alpha_{\max}^{2n} \tau'' W \right| \ge W\tau_{\max},
\end{equation}
whenever $m' \ne m$, where $S \triangleq \lbrace \left(\tau'', \alpha'' \right): \tau'' \in \left(-\tau_{\max},\tau_{\max}\right), \alpha'' \in \left(\alpha_{\max}^{-1} , \alpha_{\max}\right) \rbrace $. The condition in \eqref{eqn:cond2Eq} places an upper bound on $W$, as we shall soon see.}

First, consider a channel without Doppler, i.e., $\alpha_{\max}=1$, in which case the condition in \eqref{eqn:cond2Eq} specializes to
\begin{equation} \label{eqn:cond2EqSimple}
\underset{ \left(\tau'', 1 \right) \in S}{\inf} \:  \left| m' - m - \tau'' W \right| \ge W\tau_{\max},
\end{equation}
whenever $m' \ne m$. The condition in \eqref{eqn:cond2EqSimple}, for a Doppler-free channel, is satisfied if we choose
\begin{equation}
W \le \frac{1}{2 \tau_{\max}}. \label{eqn:wZeroDopl}
\end{equation}
This implies that the duration of the transmitted signal, $s(t)$, must be larger than $2 M \tau_{\max}$ in a Doppler-free channel having a delay spread of $2\tau_{\max}$. The choices $W = \frac{1}{2 \tau_{\max}}$ and $q=1$ for a Doppler-free channel, and the robust bi-orthogonality property, render the cross ambiguity $A_{g_{\text{rx}},g_{\text{tx}}}(\tau, 1)=0$, $\tau \notin \left(-\tau_{\max} , \tau_{\max}\right)$. Notice that, for a Doppler-free channel, with the choice of $q=1$ we must use $N=1$ and the ODSS modulation scheme defaults to asymmetric OFDM (A-OFDM), which is a scheme that converts delay-spread channels into a single tap complex channel in the Fourier domain. This behavior is very similar to the OTFS modulation scheme \cite{RavitejaJune2019}.

Finally, we discuss the choice of $W$ in ODSS modulation for a doubly-spread delay-scale channel that is both delay-spread and Doppler-distorted. 
%
Let $m' > m$. Now, if the condition in \eqref{eqn:cond2Eq} is satisfied by $m' = m+1$, then it will be satisfied by every $m ' > m$. The expression $\left| \alpha'' \alpha_{\max}^{2n} m' - m - \alpha'' \alpha_{\max}^{2n} \tau'' W \right|$, in \eqref{eqn:cond2Eq}, is minimized by $\alpha'' = \alpha_{\max}^{-1}$ and $\tau'' = \tau_{\max}$, when $W$ is such that $W \tau_{\max} < 1$, and hence $\alpha'' \alpha_{\max}^{2n} m' - m - \alpha'' \alpha_{\max}^{2n}  \tau'' W > 0$. For these settings, with $m' = m+1$, we find that $\left| \alpha'' \alpha_{\max}^{2n} m' - m - \alpha'' \alpha_{\max}^{2n} \tau'' W \right| \ge W \tau_{\max} \implies  \alpha_{\max}^{2n-1} + (\alpha_{\max}^{2n-1}-1)m - \alpha_{\max}^{2n-1} \tau_{\max} W \ge W \tau_{\max}$ and hence 
\begin{equation}
W \le \frac{ \alpha_{\max}^{2n-1} + (\alpha_{\max}^{2n-1}-1)m}{ \left( 1 + \alpha_{\max}^{2n-1}\right) \tau_{\max} }.
\end{equation}
Therefore, if 
\begin{equation}\label{eqn:wCond1}
W \le \frac{1}{ \left(1+ \alpha_{\max}\right) \tau_{\max} } \triangleq W_{m'>m},
\end{equation}
the condition in \eqref{eqn:cond2Eq} is satisfied for $m'>m$. A similar argument, for $m'<m$, leads us to the following bound on~$W$ for satisfying the condition in \eqref{eqn:cond2Eq}:
\begin{equation}\label{eqn:wCond2}
W \le \frac{1}{ \left( 1 + \alpha_{\max}^{2N-3}\right) \tau_{\max} } \triangleq W_{m'<m}.
\end{equation}
To satisfy both \eqref{eqn:wCond1} and \eqref{eqn:wCond2}, for every $m' \ne m$, we choose 
\begin{equation}\label{eqn:wChoice}
W=\min \left( W_{m'>m}, W_{m'<m}. \right) 
\end{equation}

Equations \eqref{eqn:qChoice} and \eqref{eqn:wChoice} provide the choices of the parameters $q$ and $W$, respectively, for the ODSS modulation. For the Doppler-free channel ($\alpha_{\max}=1$), we observe that the choice of $W$ reduces to $W = \frac{1}{2\tau_{\max}}$, which agrees with \eqref{eqn:wZeroDopl}. In a Doppler-distorted channel, we see that the choice of $W$ according to \eqref{eqn:wChoice} entails a longer transmit signal duration compared to a Doppler-free channel. 

\textcolor{black}{\subsubsection{The ODSS input-output relation} 
The ODSS demodulator output is obtained by taking the discrete Mellin-Fourier transform of the delay-scale signal in \eqref{eqn:odssDWTOutput}:}
\begin{multline} 
\hat{y}[k,l] =  \sum_{n = 0}^{N-1} \sum_{m = 0}^{M(n)-1} q^{n/2} \hat{Y}[n,m] e^{j 2 \pi \left(  \frac{nk}{N} - \frac{ml}{M(n)} \right)}, \\
= \sum_{n = 0}^{N-1} \sum_{m = 0}^{M(n)-1} h_w[l-m, k-n] \: x[n,m]  + w[k,l],  \label{eqn:odssIO}
\end{multline}
where $h_w[l,k]$ is obtained by sampling the frequency $\left(f=\frac{lW}{M}\right)$ and Mellin variable $\left(\beta =\frac{k}{N \ln q}\right)$ arguments of the function
\begin{multline}  \label{eqn:hwODSS}
h_w(f, \beta) = \sum_{n=0}^{N-1} \sum_{m=0}^{M(n)-1} H_{n,m}[n,m] e^{j2\pi \left(  \beta n \ln q - \frac{m}{W} f \right)}  \\
=  \iint  h(\tau',\alpha')  \sum_{m=0}^{M-1} \sum_{n=0}^{N-1} e^{j2\pi \left(  \beta n \ln q  - f \frac{m}{W} \right)}  \\ 
\: \times A_{g_{\text{rx}},g_{\text{tx}}}\left( q^{n} \left( \frac{m}{ W}\left(\alpha'-1\right) - \alpha' \tau' \right),\frac{1}{\alpha'}\right) d\tau' d\alpha'. 
\end{multline}

\emph{An Example:} Consider the channel response due to a collection of discrete reflectors associated with path delays and Doppler scales $(\tau_i, \alpha_i), i=1,2,\ldots,P$:
\begin{equation}  \label{eqn:pathModel}
h(\tau,\alpha) = \sum_{i=1}^{P} h_i \delta(\tau-\tau_i) \delta(\alpha-\alpha_i).
\end{equation}

For the above channel response model, we find that $H_{n,m}[n,m]$ in \eqref{eqn:Hnm0_} evaluates to
\begin{multline}  \label{eqn:Hnm_DS_Discrete}
H_{n_,m}[n, m] = \sum_{i=1}^{P} h_i  \\ 
\times A_{g_{\text{rx}},g_{\text{tx}}}\left( q^{n} \left( \frac{m}{ W}\left(\alpha_i-1\right) - \alpha_i \tau_i \right),\frac{1}{\alpha_i}\right), 
\end{multline}
and hence 
\begin{multline}  \label{eqn:hwODSSdiscrete}
h_w[l,k] = \sum_{i=1}^{P}  h_i  \sum_{n'=0}^{N-1} \sum_{m'=0}^{M(n')-1}  e^{j2\pi \left(  \frac{n'k}{N}  - \frac{m'l}{M(n')} \right)} \\
\times A_{g_{\text{rx}},g_{\text{tx}}}\left( q^{n'} \left( \frac{m'\left(\alpha_i-1\right)}{ W} - \alpha_i \tau_i \right),\frac{1}{\alpha_i}\right).
\end{multline}

The input-output relation in an ODSS system, given by  \eqref{eqn:odssIO}, can be depicted in the following vectorized form:
\begin{equation}  \label{eqn:odssIO_vector}
\mathbf{y} = \mathbf{H} \mathbf{x} + \mathbf{w},
\end{equation}
where $\mathbf{y} \in \mathbb{C}^{M_{\text{tot}} \times 1}$ is the output of the ODSS demodulator obtained by stacking $y[k,l]$ into a vector,  $\mathbf{H} \in \mathbb{C}^{M_{\text{tot}} \times M_{\text{tot}}}$ is the effective channel matrix, $\mathbf{x} \in \mathbb{C}^{M_{\text{tot}} \times 1}$ is the symbol vector obtained by stacking $x[k,l]$ into a vector, and $\mathbf{w} \in \mathbb{C}^{M_{\text{tot}} \times 1}$ is the additive noise at the ODSS demodulator output. 

\subsubsection{Data Decoding}
Using \eqref{eqn:odssIO_vector}, we can  use either an MMSE decoder or a message passing based decoder to recover the transmitted data symbols. This involves equalizing a channel matrix of size $M_{\text{tot}} \times M_{\text{tot}}$ which is not close to diagonal. As a consequence, it is  computationally expensive to perform channel equalization in the Mellin-Fourier domain signal represented by~\eqref{eqn:odssIO_vector}. 

Motivated by the above, we now present an alternative, simple, decoder that uses a subcarrier-by-subcarrier MMSE equalizer in the delay-scale domain instead of the Mellin-Fourier domain. We first express \eqref{eqn:odssDWTOutput} in a matrix-vector form as follows:
\begin{equation}  \label{eqn:odssDWTOutputvecForm}
\mathbf{\hat{Y}} = \mathbf{D} \mathbf{X} + \mathbf{W},
\end{equation}
where $\mathbf{\hat{Y}} \in \mathbb{C}^{M_{\text{tot}} \times 1}$ is a vector obtained by stacking the outputs $Y[n,m]$, $\mathbf{D}  \in \mathbb{C}^{M_{\text{tot}} \times M_{\text{tot}}}$ is a diagonal matrix formed by stacking $H_{n,m}[n, m]$ along its diagonal, $\mathbf{X} \in \mathbb{C}^{M_{\text{tot}} \times 1}$ contains  the data symbols obtained by stacking $X[n,m]$, and $\mathbf{W} \in \mathbb{C}^{M_{\text{tot}} \times 1}$ is the additive noise. 

\textcolor{black}{Data decoding proceeds after an MMSE equalizer on $\mathbf{\hat{Y}}$:
\begin{equation}  \label{eqn:ODSSmmse}
\mathbf{\hat{Z}} = \mathbf{D} ^H \left(\mathbf{D} \mathbf{D}^H + \sigma_W^2 I \right)^{-1} \mathbf{\hat{Y}},
\end{equation}		
where $\sigma_W^2$ is the noise variance in the delay-scale domain.}

The data symbol vector is then obtained as follows:
\begin{equation}  \label{eqn:odssDecoder}
\mathbf{\hat{x}} = \mathcal{S} \left( \mathcal{T}_{\text{iMF}}^{-1} \mathbf{\hat{Z}} \right),
\end{equation}
where the operator $\mathcal{S}(.)$ slices each entry in the input vector to the nearest symbol in the transmitted constellation.

\textcolor{black}{\emph{Remarks:} If the robust bi-orthogonality condition is not satisfied exactly, as in the case of OTFS, expression \eqref{eqn:odssIo} will not reduce to \eqref{eqn:odssIo_} and hence the measurement model \eqref{eqn:odssDWTOutput} in the delay-scale domain will not ensue. In that case, $\mathbf{D}$ will not be exactly diagonal. The matrix $\mathbf{D}$ will be \emph{nearly} diagonal if the robust bi-orthogonality is approximately satisfied (see \textcolor{black}{\cite[Sec. \ref{supp:plots}, Fig. \ref{fig:odssDelayScaleDomainChanMat}]{p2021orthogonal}}), so that it is reasonable to consider a diagonal approximation to $\mathbf{D}$ for equalization purposes. Such an approximation cannot be made in the Mellin-Fourier domain because, even if $\mathbf{D}$ is near-diagonal, $\mathbf{H}$ is not close to diagonal (see \textcolor{black}{\cite[Sec. \ref{supp:plots}, Fig. \ref{fig:odssMellinFourierDomainChanMat}]{p2021orthogonal}}).} 

\textcolor{black}{\subsubsection{Computational Complexity} The ODSS transmitter implements transformation from Mellin-Fourier domain to the delay-scale domain at the transmitter followed by the modulator that generates the waveform to transmit. The $M_{\text{tot}} \times  M_{\text{tot}}$  transform matrix is a fixed precoder matrix that can be precomputed and stored in the memory. The computational complexity of the transmitter is, therefore, $\mathcal{O}(M_{\text{tot}}^2)$. The ODSS receiver performs matched filtering, subcarrier-by-subcarrier equalization and inverse Mellin-Fourier transformation. Assuming that matched filtering is performed in the receiver front-end and that the inverse transform matrix is precomputed, the computational complexity of the receiver is also $\mathcal{O}(M_{\text{tot}}^2)$, excluding channel estimation overheads. For the same symbol rate, OFDM and OTFS have a lower complexity of $\mathcal{O}(M_{\text{tot}} \log M_{\text{tot}})$ due to efficient computations based on the Fast Fourier Transform (FFT) algorithm. The relation between Mellin and the Wavelet/Fourier transforms can be exploited to speed up the transform computations in ODSS also \cite{Antonio}. Development of a computationally efficient architecture for ODSS is beyond the scope of this paper.}

\section{Transmit and Receive Filters}{\label{sec:pulseShapeFn}}

The transmit and receive filters (pulse shaping functions), $g_{\text{tx}}(t)$ and $g_{\text{rx}}(t)$,  are required to be bi-orthogonal in a robust manner, as described in sections \ref{sec:OTFS} and \ref{sec:ODSS}, for both OTFS and ODSS modulations. However,  this is not possible (see~\textcolor{black}{\cite[Sec. \ref{supp:proof_OTFS_NonBiorthogonality} and Sec. \ref{supp:proof_ODSS_NonBiorthogonality}]{p2021orthogonal}}): we cannot find transmit and receive filters that exactly satisfy robust bi-orthogonality. Consequently, in most implementations of OTFS, pulse shaping functions such as rectangular, raised cosine and Dolph-Chebyshev windows are employed both at the transmitter and receiver~\cite{Raviteja2018}. We adopt a similar  approach in ODSS and show its effectiveness through numerical simulations; in particular, our choice of ODSS subcarriers results in the channel matrix $D$ represented in \eqref{eqn:odssDWTOutputvecForm} becoming nearly diagonal in the delay-scale domain.  

\textcolor{black}{In this work,  to form the transmit pulse-shaping filter, we employ a basic chirplet generated by linearly sweeping frequency from $f_1 = \frac{1}{\sqrt{q}}$ to $f_2 = \sqrt{q}$ in $T$ seconds:
\begin{equation} \label{eqn:gChip}
g_{\text{0}}(t) = e^{j 2 \pi \left(f_1 t + \frac{1}{2} \kappa t^2\right)}, 0 \le t \le T,
\end{equation}
where $\kappa = \frac{f_2 - f_1}{T}$ is the chirp sweep rate. The basic chirplet duration $T$ is also the ODSS symbol duration.} 

\textcolor{black}{To reduce the spectral sidelobes, we apply a PHYDYAS filter based window \cite{Bellanger2010} to obtain the transmit pulse-shaping filter  $g_{\text{tx}}(t) =  g_{\text{w}}(t) g_{\text{0}}(t)$}, 
where $g_{\text{w}}(t)$ is the window function given by 
\begin{equation} \label{eqn:gW}
g_{\text{w}}(t) =  1 + 2 \sum_{k=1}^{K-1} (-1)^k A[k] \cos \left( \frac{2\pi k t}{KT} \right), 
\end{equation}
with $A[k], k=1,2,\ldots, K-1,$ being the PHYDYAS reference filter coefficients. \textcolor{black}{In simulations, we use an overlap factor of $K=3$, for which the PHYDYAS filter coefficient values are: $A[1] = 0.91143783$ and $A[2]= 0.41143783$ \cite{Bellanger2010}.}

\textcolor{black}{From the above linearly modulated pulse, or \emph{chirplet}, the ODSS subcarrier waveforms are generated by $q$-adic compression and shifting (see \eqref{eqn:odssModulator}):
\begin{equation} \label{eqn:odssSubcarriers}
s_{m,n}(t) = q^{n/2} g_{\text{tx}} \left(q^n \left(t-\frac{m}{q^nW}\right)\right), 
\end{equation}	
where $m = 0, 1, \ldots, M(n)-1$, $M(n) = \lfloor q^n\rfloor$, and $n = 0, 1, \ldots, N - 1$. 
We will see that these subcarriers, when used over a delay-scale spread channel, result in a sparse and nearly diagonal channel matrix \textcolor{black}{in the delay-scale domain, enabling the use of low complexity receivers in that domain.}}

\section{Numerical Results}{\label{sec:numResults}}

In this section, we investigate the bit error rate (BER) performance of the  ODSS modulation scheme. To that end, we first design the subcarriers of ODSS modulation respecting the criteria developed in Sec. \ref{sec:ODSS} to avoid ICI. Although robust bi-orthogonality cannot be satisfied in an exact manner (the same as in the case of OTFS) due to the reasons mentioned in Sec. \ref{sec:pulseShapeFn}, we use transmit and receive pulses that lead to a low complexity receiver. In subsection \ref{subsec:odssDesign}, we discuss these aspects. In the  subsection \ref{subsec:berPerf}, we present BER performance results for the ODSS modulation scheme designed in subsection \ref{subsec:odssDesign}.

\subsection{ODSS Waveform }{\label{subsec:odssDesign}}

Recall our discussions in Sec. \ref{subsec:odssReceiver} leading to the choice of geometric sampling ratio, $q$, and the transmit filter bandwidth, $W(q)$, for avoiding ICI in ODSS modulation. From the discussion preceding \eqref{eqn:qChoice}, we choose $q$ such that $ \sqrt{q} > \alpha_{\max} \ge 1$. 


Let $B$ denote the system bandwidth and $N$ be the number of $q$-adic scales (compressions) on the scale axis. 
\textcolor{black}{The bandwidth occupied by the transmit pulse shaping filter and all its time-compressed copies is $\sum_{n=0}^{N-1} q^n W(q)$.} 
Clearly, for the transmit signals to fit within the system bandwidth, we need:
 \begin{equation}\label{eqn:txFiltBW}
 	\sum_{n=0}^{N-1} q^n W(q) < B.
 \end{equation}
 
 The maximum allowable transmit filter bandwidth is given by (see \eqref{eqn:wChoice})
 \begin{equation} \label{eqn:Wmax}
 W_{\max}(q,N) \triangleq \begin{cases}
 \frac{1}{ \left( 1 + \alpha_{\max} \right) \tau_{\max} },  & N=1, \\
 \frac{1}{ \left( 1 +  \alpha_{\max}^{2N-3}\right) \tau_{\max} }, & \text{otherwise}.
 \end{cases}	
 \end{equation}
 
 Considering \eqref{eqn:txFiltBW}, the upper limit (in \eqref{eqn:Wmax}) on the transmit filter bandwidth is satisfied if:
  \begin{equation}\label{eqn:searchN}
W(q) < \frac{B}{\sum_{n=0}^{N-1} q^n} = \frac{B(q-1)}{q^{N}-1}< W_{\max}(q,N).
 \end{equation}
 
 We choose the number of $q$-adic scales, to be the smallest integer $N$, say, $N(q)$, that satisfies \eqref{eqn:searchN}. We choose the transmit filter bandwidth, $W(q)$, to be
   \begin{equation}\label{eqn:txFiltBW_q}
 W(q) = \frac{B(q-1)}{q^{N(q)}-1}.
 \end{equation}
Then, the number of symbols that can be mounted is $M_{\text{tot}}(q) = \sum_{n=0}^{N(q)-1} \lfloor q^n \rfloor$.

Let the duration of the ODSS symbol block be $T(q) = \frac{\gamma}{W(q)}$, where $\gamma>1$  a factor that accounts for the increase in length of the filter above the minimum duration of $\frac{1}{W(q)}$. Note that the choice of $N = N(q)$ results in the smallest ODSS symbol duration that can be used. Then, the spectral efficiency of the ODSS modulation scheme (in symbols/s/Hz) is given by
	\begin{equation}\label{eqn:eta}
		\eta(q) = \frac{M_{\text{tot}}(q)}{BT(q)} = \frac{M_{\text{tot}}(q)W(q)}{\gamma B} = \frac{M_{\text{tot}}(q)(q-1)}{\gamma (q^{N(q)}-1)}.
	\end{equation}
	
\textcolor{black}{In \cite[Sec. \ref{supp:specEff}]{p2021orthogonal}, through a numerical example, we demonstrate that ODSS can operate with a spectral efficiency close to one symbol per second per Hertz.}



\begin{figure}
		\setxysizeo
		\centering
		\includegraphics[scale=.42]{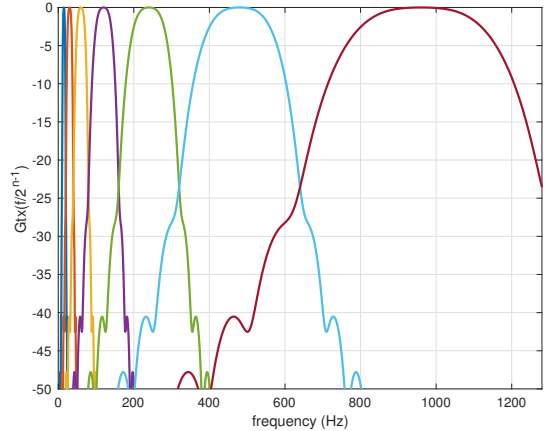}
		\caption{ODSS subcarrier spectra. Note that the seven subcarriers, for $n=0,1,\ldots,6$, span a frequency band of $0-1280$~Hz.}\label{fig:odssSubcarrierSpectra}
		\pullUp
		\vspace{-0.3cm}
	\end{figure}

Consider the ODSS subcarrier waveforms, constructed as discussed in Sec. \ref{sec:pulseShapeFn}, on a dyadic ($q=2$) tiling in the delay-scale space for a symbol block duration of $T=1.9$ seconds and time-scale indices $n=0,1,\ldots,6$. Fig.~\ref{fig:odssSubcarrierSpectra} shows the ODSS subcarrier spectra. From these figures, we notice that the subcarrier bandwidth doubles for every scale increment and so does the number of time-compressed and shifted subcarriers at each scale. \textcolor{black}{The ODSS waveforms, thus constructed, are nearly orthogonal (see \cite[Sec. \ref{supp:specEff}]{p2021orthogonal} for more details).}


\subsection{BER Performance}{\label{subsec:berPerf}}

We turn to investigate the communication performance of the  ODSS scheme. 
We simulate the transmitter and receiver of three schemes -- OFDM, OTFS and ODSS -- operating in a doubly-spread (i.e., time scale  and delay spread) channel. We evaluate the bit error rate (BER) performance as the signal to noise ratio (SNR) at the receiver is varied. \textcolor{black}{For fair comparison, we evaluate all three schemes with a low-complexity subcarrier-by-subcarrier MMSE equalizer at the receiver. In the case of OFDM, channel equalization is performed in the frequency domain where the symbols are mounted. Subcarrier-by-subcarrier MMSE channel equalizer is implemented in the time-frequency (resp. delay-Doppler) domain outputs for OTFS (resp. ODSS). The evaluation of the performance with more computationally expensive message passing based equalizers is relegated to future work.}

We define the SNR as the ratio of the signal and noise powers at the receiver front-end. The transmitted ODSS signal waveform, given in \eqref{eqn:odssModulator}, can be expressed as 
\begin{equation}  \label{eqn:odssSignal}
	\mathbf{s} = \mathbf{G} \mathbf{X},
\end{equation}
where the columns of the matrix $\mathbf{G}$ are the basis waveforms (compressed and shifted versions of chirplets), $\mathbf{X} = \mathcal{T}_{\text{iMF}} \mathbf{x}$, \textcolor{black}{$\mathbf{x}$ being the vectorized version of the symbols on the 2D-Mellin-Fourier domain grid.} 
The signal at the receiver, after the transmitted ODSS waveform propagates through a doubly-spread channel in \eqref{eqn:pathModel}, is given by: $r_s(t) = \sum_{p=1}^{P} h_p s \left( \alpha_p (t- \tau_p) \right)$. 

To compute the SNR at the receiver side, we ignore the effect of time scale ($0.999 < \alpha_p < 1.001$) since we need only the power of the signal component in the receiver waveform. The received signal power is given by
\begin{equation}  \label{eqn:odssSignalRxPow}
	P_s = \mathbb{E}(|r_s(t)|^2) =  \mathbb{E}( \sum_{i=1}^{P} \sum_{j=1}^{P} h_i^* h_j s^*(t- \tau_i) s(t - \tau_j) ). 
\end{equation}

Next, assuming that the channel coefficients $\lbrace h_p \sim \mathcal{CN}(0,1): p = 1, 2, \ldots, P \rbrace$ are mutually independent and independent of the transmitted signal, we find
\begin{equation}  \label{eqn:odssSignalRxPow_2}
	P_s =\sum_{i=1}^{P} \mathbb{E}(|h_i|^2) \mathbb{E}{|s(t - \tau_i)|^2} = P \mathbb{E}{|s(t)|^2},
\end{equation}
where we made use of the fact that power of the signal is not affected by delay. Therefore, making use of \eqref{eqn:odssSignal}, we have
\begin{equation}  \label{eqn:odssSignalRxPow_3}
	P_s = \mathbb{E}{\lbrace \mathbf{X}^H \mathbf{G}^H \mathbf{G} \mathbf{X}\rbrace} = \frac{1}{F_sT} \text{Tr}\lbrace   \mathbf{G} \mathbb{E}\left[ \mathbf{X} \mathbf{X}^H \right] \mathbf{G}^H \rbrace,
\end{equation}
where $F_s$ is the sampling rate. Since the ODSS transform preserves energy, $\mathbb{E}\left[ \mathbf{X} \mathbf{X}^H \right] = \mathbb{E}\left[ \mathbf{x} \mathbf{x}^H \right] = \mathbf{I}$ and therefore
\begin{equation}  \label{eqn:odssSignalRxPow_4}
	P_s =  \frac{1}{F_sT} \text{Tr}\lbrace   \mathbf{G} \mathbf{G}^H \rbrace.
\end{equation}

	\begin{figure}[t]
	\setxysizeo
	\centering
	\includegraphics[scale=.42]{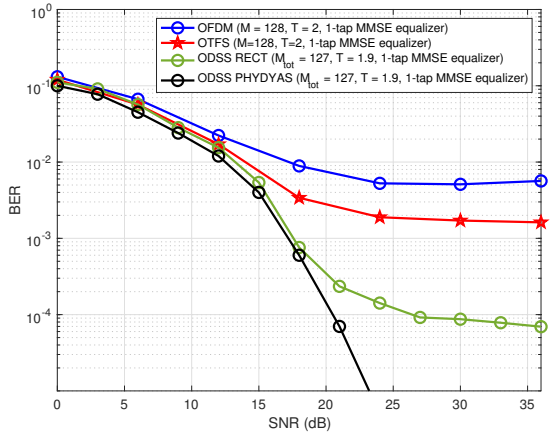}
	\caption{BER performance of OFDM, OTFS and ODSS schemes using one-tap MMSE channel equalizers in a wideband delay-scale spread channel with $\tau_{\max} = 10$~ms, $\alpha_{\max} = 1.001$ and $P = 20$ paths.}	
	\label{fig:BERvSNRperformance}
	\pullUp
	\vspace{-0.3cm}
	\end{figure}

\begin{figure}[h]
	\setxysizeo
	\centering
	\includegraphics[scale=.42]{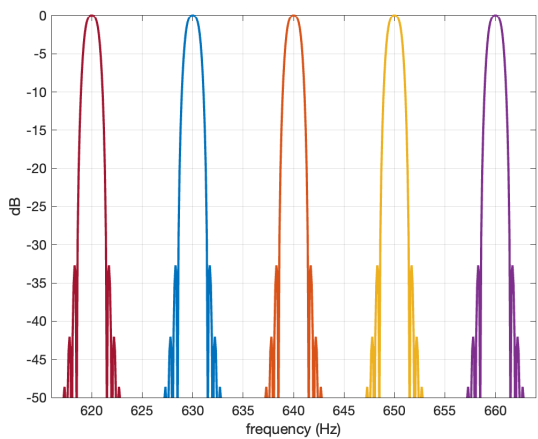}
	\caption{Spectrum of the pulse shaped OFDM subcarriers in the frequency band 615-665 Hz. The subcarriers are spaced at $\Delta F = 10$ Hz. The 3-dB spectral width of the subcarriers, after pulse shaping by PHYDYAS filter, is $W=1.5$ Hz. }\label{fig:ofdmSubcarrierSpectra}
	\pullUp
	\vspace{-0.2cm}
	\end{figure}

We compare the three schemes when they operate at the same spectral efficiency. For BER performance evaluation, we consider acoustic communications in the frequency band, $[f_c - B/2, f_c + B/2]$, where $f_c = 12.8$ kHz and $B = 1.28$ kHz. Both OFDM and OTFS use $N_{\text{FFT}}=2560$ point FFT and they mount every twentieth subcarrier with a binary phase shift keying (BPSK) symbol. The receivers first down-convert the received signal to the frequency band from DC to $1280$~Hz. Then, they use a waveform sampling rate of $F_s = 1280$ Hz and a PHYDYAS filter (with an overlap factor $K=3$) for pulse shaping. The spectral width of the subcarriers has increased by three-fold, from $W=0.5$ Hz to $W=1.5$ Hz, due to pulse shaping by PHYDYAS filter. The utilized subcarriers of the OFDM are spaced well apart, by $\Delta F = 10$ Hz, with significant guard band and without overlap (see \textcolor{black}{\cite[Sec. \ref{supp:plots}, Fig. \ref{fig:ofdmSubcarrierSpectra}]{p2021orthogonal}}). 

\textcolor{black}{Figure \ref{fig:BERvSNRperformance} shows the performance of the three modulation schemes as a function of the SNR, in a doubly-spread channel with a delay spread of $\tau_{\max} = 10$ ms, maximum Doppler scale $\alpha_{\max} = 1.001$ and number of paths $P = 20$. The path amplitudes are Rayleigh distributed, $h_p \overset{i.i.d.}{\sim} \mathcal{CN}(0,1): p=1,2, \ldots, P$. The path delays, $\tau_p$, and time-scales, $\alpha_p$, are drawn uniformly from  $(0, \tau_{\max})$ and $(1/\alpha_{\max}, \alpha_{\max})$, respectively. Notice that the delays, $\tau_p$, and time-scales, $\alpha_p$, are drawn from continuous distributions and do not necessarily lie on the sampling grid. We oversample the transmitted signal by a factor of $8$, round-off the channel delay taps to this higher-rate sampled time grid, perform resampling by a rational approximation of the resampling rates $\alpha_p$ (to within an error of $\epsilon =  10^{-5}$), obtain the received signal after propagating through the delay-scale channel, and finally downsample to obtain the received signal at the original sampling rate.} All the receivers use a low-complexity \textcolor{black}{subcarrier-by-subcarrier MMSE equalizer} based symbol decoder. We notice the superior performance of the proposed ODSS scheme, while OFDM performs the worst. \textcolor{black}{This is mainly due to the larger ICI among the high frequency subcarriers of OFDM and OTFS when compared to ODSS (see \cite[Sec. \ref{supp:ICI_ofdm_otfs}]{p2021orthogonal})}. In a time-scale channel, the frequency shift is non-uniform and increases with frequency. The ODSS subcarriers have a bandwidth that also increases with frequency, and are therefore relatively unaffected by the Doppler due to time scaling. \textcolor{black}{Fig. \ref{fig:BERvSNRperformance} also shows the performance of ODSS with both the rectangular and PHYDYAS pulse shaping filter. We see that the PHYDYAS filter reduces ICI and thereby eliminates  the error floor within the range of SNR considered.}

\begin{figure}[t]
	\setxysizeo
	\centering
	\includegraphics[scale=.42]{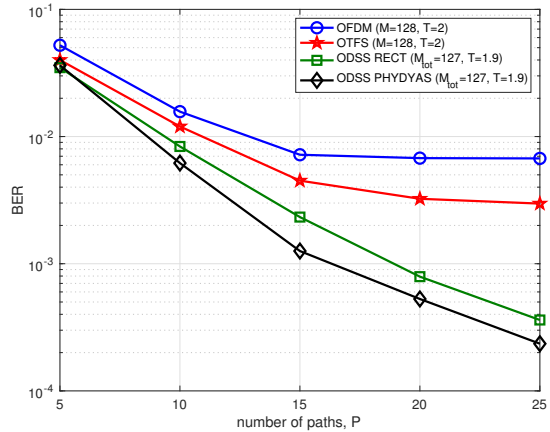}
	\caption{BER performance of OFDM, OTFS and ODSS using one-tap MMSE channel equalizer as the number of paths, $P$, is varied in a doubly-spread wideband channel with $\tau_{\max} = 10$~ms, $\alpha_{\max} = 1.001$ and at SNR$= 18$~dB. 
	}	
	\label{fig:BERvNperformance}
	\pullUp
	\vspace{-0.3cm}
	\end{figure}

Figure \ref{fig:BERvNperformance} shows the performance of the three schemes at an SNR of $18$ dB in a doubly-spread channel with a delay spread of $\tau_{\max} = 10$ ms and maximum Doppler scale $\alpha_{\max} = 1.001$, as the number of paths $P$ is varied. The performance advantage of the ODSS scheme increases with the number of paths. In a doubly-spread wideband channel, ODSS whose performance is not limited by a BER floor, unlike the other two schemes, benefits due to diversity gain as the number of paths increases.

\begin{figure}[t]
	\setxysizeo
	\centering
	\includegraphics[scale=.42]{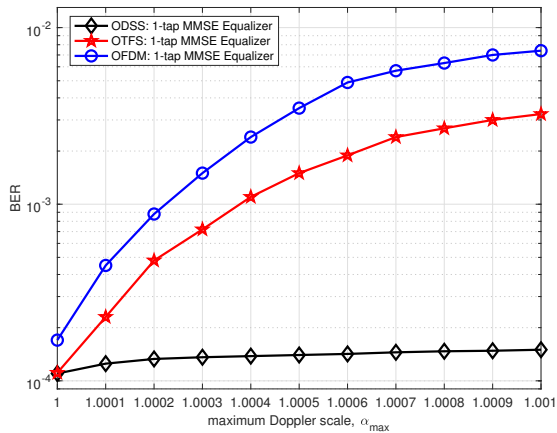}
	\caption{BER performance of OFDM, OTFS and ODSS as the Doppler scale spread parameter, $\alpha_{\max}$, is varied in a wideband channel with $P = 20$, $\tau_{\max} = 10$~ms and at SNR$= 20$~dB.}	
	\label{fig:BERvAlphaperformance}
	\pullUp
	\vspace{-0.3cm}
	\end{figure}

\textcolor{black}{Finally, we vary the maximum Doppler scale spread, $\alpha_{\max}$, from $\alpha_{\max}=1.0$ (zero Doppler channel) to $\alpha_{\max} = 1.001$ keeping all other parameters fixed ($\tau_{\max} = 10$ ms, $P=20$ and SNR = $20$ dB.). Figure \ref{fig:BERvAlphaperformance} shows the BER performance of the three schemes. The ODSS modulation, designed to handle a maximum Doppler scale spread of $\alpha_{\max} = 1.001$, has a nearly constant BER for $1.0 \le \alpha_{\max} \le 1.001$. OFDM and OTFS schemes suffer due to ICI from Doppler distortion, for $\alpha_{\max} > 1$, that gets severe as the Doppler spread increases.}

\section{Conclusions}{\label{sec:conclusions}}

We developed a new low complexity modulation scheme for a delay and Doppler time-scale spread wideband channel, which we called
Orthogonal Delay Scale Space (ODSS) modulation. We examined the performance of OFDM, OTFS and ODSS modulation schemes through numerical simulations when the receiver employs a low complexity channel equalizer. In doubly distorted wideband channels, the ODSS receiver using a \textcolor{black}{subcarrier-by-subcarrier equalizer} showed a clear performance advantage over the OFDM and OTFS receivers. Also, as the number of multipaths increased, ODSS showed even better performance, taking advantage of the increased multipath diversity, whereas the other two schemes suffered due to their inability to handle ICI.

The ODSS scheme was developed by systematically  identifying the transmitter and receiver side modulation and demodulation functions suited for \emph{wideband} time-varying channels. In the process, we introduced the 2D ODSS transform composed of the inverse Fourier and Mellin transforms from the Fourier-Mellin domain (symbol space) to the delay-scale domain (waveform space). We recognized the $\omega$-convolution property that helped in developing the input-output model for the ODSS scheme. We showed that the proposed scheme can operate with a spectral efficiency close to one symbol per second per Hertz. With our choice of filtered wideband chirplet, as the basic ODSS waveform, we obtained a channel matrix which was nearly diagonal thus allowing the use of a \textcolor{black}{subcarrier-by-subcarrier equalizer} in the delay-scale domain.


While this work introduced the ODSS modulation scheme, there are several directions that can be explored. These include developing better-performing message-passing based receivers, analyzing the peak-to-average power ratio and energy efficiency, extension to multiple antennas at the transceivers, analyzing the diversity-multiplexing gain trade-offs, \textcolor{black}{developing channel estimation schemes, analyzing performance under imperfect channel state information, designing good transmit and receive pulse shaping functions}, and so on.

\bibliographystyle{IEEEtran}
\bibliography{IEEEabrv,bibJournalList,myRefs}

\begin{thebibliography}{10}
\providecommand{\url}[1]{#1}
\csname url@samestyle\endcsname
\providecommand{\newblock}{\relax}
\providecommand{\bibinfo}[2]{#2}
\providecommand{\BIBentrySTDinterwordspacing}{\spaceskip=0pt\relax}
\providecommand{\BIBentryALTinterwordstretchfactor}{4}
\providecommand{\BIBentryALTinterwordspacing}{\spaceskip=\fontdimen2\font plus
\BIBentryALTinterwordstretchfactor\fontdimen3\font minus
  \fontdimen4\font\relax}
\providecommand{\BIBforeignlanguage}[2]{{%
\expandafter\ifx\csname l@#1\endcsname\relax
\typeout{** WARNING: IEEEtran.bst: No hyphenation pattern has been}%
\typeout{** loaded for the language `#1'. Using the pattern for}%
\typeout{** the default language instead.}%
\else
\language=\csname l@#1\endcsname
\fi
#2}}
\providecommand{\BIBdecl}{\relax}
\BIBdecl

\bibitem{HadaniMar2017}
R.~Hadani, S.~Rakib, M.~Tsatsanis, A.~Monk, A.~Goldsmith, A.~Molisch, and
  R.~Calderbank, ``Orthogonal time frequency space modulation,'' in
  \emph{{IEEE} WCNC, San Fransisco, CA}, Mar. 2017, pp. 1--6.

\bibitem{Hadani2018Orthogonal}
\BIBentryALTinterwordspacing
R.~Hadani, S.~Rakib, S.~Kons, M.~Tsatsanis, A.~Monk, C.~Ibars, J.~Delfeld,
  Y.~Hebron, A.~J. Goldsmith, A.~F. Molisch, and R.~Calderbank, ``Orthogonal
  time frequency space modulation,'' Aug. 2018. [Online]. Available:
  \url{\url{https://arxiv.org/abs/1808.00519v1}}
\BIBentrySTDinterwordspacing

\bibitem{Hadani2018otfs}
\BIBentryALTinterwordspacing
R.~Hadani and A.~Monk, ``{OTFS}: A new generation of modulation addressing the
  challenges of {5G},'' Feb 2018. [Online]. Available:
  \url{\url{https://arxiv.org/abs/1802.02623v1}}
\BIBentrySTDinterwordspacing

\bibitem{HadaniIMS2017}
R.~Hadani, S.~Rakib, A.~Molisch, C.~Ibars, A.~Monk, M.~Tsatsanis, J.~Delfeld,
  A.~Goldsmith, and R.~Calderbank, ``Orthogonal time frequency space ({OTFS})
  modulation for millimeter-wave communications systems,'' in \emph{IEEE/MTT-S
  International Microwave Symposium}, June 2017, pp. 681--683.

\bibitem{RavitejaJune2019}
P.~Raviteja, E.~Viterbo, and Y.~Hong, ``{OTFS} performance on static multipath
  channels,'' \emph{IEEE Wireless Communications Letters}, vol.~8, pp.
  745--748, June 2019.

\bibitem{Murali2018}
K.~Murali and A.~Chockalingam, ``On {OTFS} modulation for high-{Doppler} fading
  channels,'' in \emph{Information Theory and Applications Workshop}, Feb.
  2018, pp. 1--10.

\bibitem{ObergJames2004}
J.~Oberg, ``Titan calling,'' \emph{IEEE Spectrum}, vol.~41, pp. 28--33, 10
  2004.

\bibitem{Franz}
F.~Hlawatsch and G.~Matz, \emph{Wireless Communications Over Rapidly
  Time-Varying Channels}, 1st~ed.\hskip 1em plus 0.5em minus 0.4em\relax USA:
  Academic Press, Elsevier, 2011.

\bibitem{Rickard}
S.~Rickard, ``Time-frequency and time-scale representations of doubly spread
  channels,'' in \emph{PhD thesis, Princeton University}, March 2005, pp.
  281--284.

\bibitem{Balan}
R.~Balan, H.~V. Poor, S.~Rickard, and S.~Verdu, ``Time-frequency and time-scale
  representations of doubly spread channels,'' in \emph{Proc. European Signal
  Processing Conf., Vienna, Austria}, September 2004, pp. 445--448.

\bibitem{Molisch2014}
A.~Molisch and F.~Tufvesson, ``Propagation channel models for next-generation
  wireless communications systems,'' \emph{IEICE Transactions on
  Communications}, vol. E97.B, pp. 2022--2034, Oct. 2014.

\bibitem{Molisch}
A.~F. Molisch, ``Ultra-wideband propagation channels,'' in \emph{Proceedings of
  the IEEE}, vol.~97, Feb. 2009, pp. 353--371.

\bibitem{Schuster}
U.~G. Schuster, \emph{Wireless Communication Over Wideband Channels}.\hskip 1em
  plus 0.5em minus 0.4em\relax Series in Communication Theory, edited by Helmut
  B{\"{o}}lcskei, 2009, vol.~4.

\bibitem{Benedetto}
M.-G. Di~Benedetto, T.~Kaiser, A.~Molisch, I.~Oppermann, C.~Politano, and
  D.~Porcino, \emph{Ultra-wideband Communication Systems: A Comprehensive
  Overview}.\hskip 1em plus 0.5em minus 0.4em\relax Hindawi Publishing
  Corporation, 01 2006.

\bibitem{Nikookar}
H.~Lu, T.~Xu, and H.~Nikookar, \emph{Cooperative Communication over Multi-scale
  and Multi-lag Wireless Channels}.\hskip 1em plus 0.5em minus 0.4em\relax
  Intech, 2012, pp. 103--126.

\bibitem{Giannakis}
L.~Yang and G.~B. Giannakis, ``Ultra-wideband communications: An idea whose
  time has come,'' in \emph{IEEE Signal Processing Mag.}, November 2004, pp.
  26--55.

\bibitem{Haimovich}
L.~Zhao and A.~Haimovich, ``Performance of ultra-wideband communications in the
  presence of interference,'' \emph{IEEE J. Select. Areas Commun.}, vol.~20,
  pp. 1684--1691, Dec. 2002.

\bibitem{Margetts}
A.~R. Margetts, ``Joint scale-lag diversity in mobile wideband
  communications,'' in \emph{PhD thesis, The Ohio State University}, 2005.

\bibitem{Tse}
T.~Jonathan, P.~Dana, and D.~Tse, ``Prediction and modeling for the
  time-evolving ultra-wideband channel,'' \emph{J. Sel. Topics Signal
  Processing}, vol.~1, no.~3, pp. 340--356, 2007.

\bibitem{Fan}
P.~Fan, E.~Panayirci, H.~V. Poor, and P.~T. Mathiopoulos, ``Special issue on
  broadband mobile communications at very high speeds,'' \emph{EURASIP Journal
  on Wireless Communications and Networking}, no. 279, Aug. 2012.

\bibitem{He}
R.~He, Z.~Zhong, b.~ai, G.~Wang, J.~Ding, and A.~Molisch, ``Measurements and
  analysis of propagation channels in high-speed railway viaducts,'' \emph{IEEE
  Transactions on Wireless Communications}, vol.~12, pp. 794--805, 02 2013.

\bibitem{Wu2016ASO}
J.~Wu and P.~Fan, ``A survey on high mobility wireless communications:
  Challenges, opportunities and solutions,'' \emph{IEEE Access}, vol.~4, pp.
  450--476, 2016.

\bibitem{Sammna}
A.~Al-Sammna, M.~Bin~Azmi, and T.~Rahman, ``Time-varying ultra-wideband channel
  modeling and prediction,'' \emph{Symmetry}, vol.~10, p. 631, Nov. 2018.

\bibitem{Davies}
J.~Davies, S.~Pointer, and S.~Dunn, ``Wideband acoustic communications
  dispelling narrowband myths,'' in \emph{MTS/IEEE OCEANS, Providence, RI,
  USA}, vol.~1, Sep. 2000, pp. 377--384.

\bibitem{Stojanovic96}
M.~Stojanovic, ``Recent advances in high-speed underwater acoustic
  communications,'' \emph{IEEE J. Oceanic. Eng.}, vol.~21, no.~2, pp. 125--136,
  Apr. 1996.

\bibitem{Stojanovic2013}
P.~Qarabaqi and M.~Stojanovic, ``Statistical characterization and
  computationally efficient modeling of a class of underwater acoustic
  communication channels,'' \emph{{IEEE} J. Ocean. Eng.}, vol.~38, no.~4, pp.
  701--717, Oct. 2012.

\bibitem{Papandreou}
Y.~Jiang and A.~{Papandreou-Suppappola}, ``Discrete time-scale characterization
  of wideband time-varying systems,'' \emph{{IEEE} Trans. Signal Process.},
  vol.~54, no.~4, pp. 1364--1375, April 2006.

\bibitem{PapandreouConf}
------, ``Time-scale canonical model for wideband system characterization,'' in
  \emph{Proc. IEEE Int. Conf. Acoustics, Speech, Signal Processing}, March
  2005, pp. 281--284.

\bibitem{Preisig}
M.~Stojanovic and J.~Preisig, ``Underwater acoustic communication channels:
  Propagation models and statistical characterization,'' \emph{{IEEE} Commun.
  Mag.}, vol.~47, no.~1, pp. 84--89, Jan. 2009.

\bibitem{Roudsari2017ChannelMF}
H.~M. Roudsari, J.~Bousquet, and G.~McIntyre, ``Channel model for wideband
  time-varying underwater acoustic systems,'' \emph{IEEE OCEANS, Aberdeen}, pp.
  1--7, 2017.

\bibitem{Zakharov}
C.~Liu, Y.~V. Zakharov, and T.~Chen, ``Doubly selective underwater acoustic
  channel model for a moving transmitter/receiver,'' \emph{IEEE Transactions on
  Vehicular Technology}, vol.~61, no.~3, pp. 938--950, 2012.

\bibitem{Fengzhong}
F.~Qu and L.~Yang, ``Basis expansion model for underwater acoustic channels?''
  in \emph{IEEE OCEANS, Quebec City, Canada}, 2008, pp. 1--7.

\bibitem{Eggen}
T.~Eggen, A.~Baggeroer, and J.~Preisig, ``Communication over {Doppler} spread
  channels. {Part I}: Channel and receiver presentation,'' \emph{IEEE Journal
  of Oceanic Engineering}, vol.~25, no.~1, pp. 62--71, 2000.

\bibitem{Dhanoa}
J.~Dhanoa, R.~Ormondroyd, and E.~Hughes, ``A robust digital communication
  system for doubly-spread underwater acoustic channels,'' in \emph{Proceedings
  of MTS/IEEE OCEANS, Kobe, Japan}, vol.~1, 2004, pp. 9--13.

\bibitem{Sadia2008}
S.~Ahmed and H.~Arslan, ``Evaluation of frequency offset and {Doppler} effect
  in terrestrial {RF} and in underwater acoustic {OFDM} systems,'' in
  \emph{IEEE MILCOM, San Diego, CA, USA}, 2008, pp. 1--7.

\bibitem{Freitag2006}
M.~Stojanovic and L.~Freitag, ``Multichannel detection for wideband underwater
  acoustic {CDMA} communications,'' \emph{IEEE J. of Ocean. Eng.}, vol.~31,
  no.~3, pp. 685--695, Jul. 2006.

\bibitem{Berger2008}
S.~F. Mason, C.~R. Berger, S.~Zhou, and P.~Willett, ``Detection,
  synchronization, and {Doppler} scale estimation with multicarrier waveforms
  in underwater acoustic communication,'' \emph{IEEE Journal on Selected Areas
  in Communications}, vol.~26, no.~9, pp. 1638--1649, 2008.

\bibitem{Berger2010SStoCS}
C.~R. Berger, S.~Zhou, J.~C. Preisig, and P.~Willett, ``Sparse channel
  estimation for multicarrier underwater acoustic communication: From subspace
  methods to compressed sensing,'' \emph{{IEEE} Trans. Signal Process.},
  vol.~57, no.~5, pp. 2941--2965, May 2011.

\bibitem{Berger2011ProgICIEq}
J.~Z. Huang, S.~Zhou, J.~Huang, C.~R. Berger, and P.~Willett, ``Progressive
  inter-carrier interference equalization for {OFDM} transmission over
  time-varying underwater acoustic channels,'' \emph{{IEEE} J. Sel. Topics
  Signal Process.}, vol.~5, no.~8, pp. 1524--1536, Dec. 2011.

\bibitem{Schlegel}
C.~Schlegel, D.~Truhachev, Z.~Alavizadeh, and A.~Vaezi, ``Effective
  communications over doubly-selective acoustic channels using iterative signal
  cancelation,'' in \emph{Proceedings of the International Conference on
  Underwater Networks \& Systems}.\hskip 1em plus 0.5em minus 0.4em\relax New
  York, NY, USA: Association for Computing Machinery, 2017.

\bibitem{Shengli2007}
B.~Li, S.~Zhou, M.~Stojanovic, L.~Freitag, and P.~Willett, ``Non-uniform
  {Doppler} compensation for zero-padded {OFDM} over fast-varying underwater
  acoustic channels,'' in \emph{IEEE OCEANS, Aberdeen, UK}, 2007, pp. 1--6.

\bibitem{Yerramalli2012PartialFFT}
S.~Yerramalli, M.~Stojanovic, and U.~Mitra, ``Partial {FFT} demodulation: A
  detection method for highly {Doppler} distorted {OFDM} systems,''
  \emph{{IEEE} Trans. Signal Process.}, vol.~60, no.~11, pp. 5906--5918, Nov.
  2012.

\bibitem{SparseComparisonHuang}
Y.~Huang, L.~Wan, S.~Zhou, Z.~H. Wang, and J.-Z. Huang, ``Comparison of sparse
  recovery algorithms for channel estimation in underwater acoustic {OFDM} with
  data-driven sparsity learning,'' \emph{Elsevier Physical Communication},
  vol.~13, no.~3, pp. 156--167, Dec. 2014.

\bibitem{Li2008Nonuniform}
B.~Li, S.~Zhou, M.~Stojanovic, L.~Freitag, and P.~Willett, ``Multicarrier
  communication over underwater acoustic channels with nonuniform {Doppler}
  shifts,'' \emph{{IEEE} J. Ocean. Eng.}, vol.~33, no.~2, pp. 1638--1649, Apr.
  2008.

\bibitem{Freitag1997}
M.~Johnson, L.~Freitag, and M.~Stojanovic, ``Improved {Doppler} tracking and
  correction for underwater acoustic communications,'' in \emph{IEEE
  International Conference on Acoustics, Speech, and Signal Processing},
  vol.~1, 1997, pp. 575--578.

\bibitem{Sibul1994}
L.~H. Sibul, L.~G. Weiss, and T.~L. Dixon, ``Characterization of stochastic
  propagation and scattering via {Gabor} and wavelet transforms,''
  \emph{Journal of Computational Acoustics}, vol.~02, no.~03, pp. 345--369,
  1994.

\bibitem{BertrandBook}
J.~Bertrand, P.~Bertrand, and J.-P. Ovarlez, \emph{{The {Mellin} Transform}},
  2nd~ed.\hskip 1em plus 0.5em minus 0.4em\relax Boca Raton: CRC Press LLC,
  2000.

\bibitem{Antonio}
A.~Sena and D.~Rocchesso, ``A fast {Mellin} and scale transform,''
  \emph{EURASIP Journal on Advances in Signal Processing}, vol. 2007, pp. 1--9,
  Jan. 2007.

\bibitem{Oppenheim}
A.~V. Oppenheim, A.~S. Willsky, and S.~H. Nawab, \emph{{Signals and Systems}},
  2nd~ed.\hskip 1em plus 0.5em minus 0.4em\relax Englewood Cliffs, N.J.:
  Prentice-Hall, Inc., 1996.

\bibitem{Bertrand1990}
J.~Bertrand, P.~Bertrand, and J.~Ovarlez, ``Discrete {Mellin} transform for
  signal analysis,'' in \emph{International Conference on Acoustics, Speech,
  and Signal Processing}, vol.~3, May 1990, pp. 1603--1606.

\bibitem{p2021orthogonal}
\BIBentryALTinterwordspacing
K.~P. Arunkumar and C.~R. Murthy, ``Orthogonal delay scale space modulation: A
  new technique for wideband time-varying channels,'' Nov. 2021. [Online].
  Available: \url{\url{https://arxiv.org/abs/2111.10765}}
\BIBentrySTDinterwordspacing

\bibitem{Raviteja2018}
P.~Raviteja, Y.~Hong, E.~Viterbo, and E.~Biglieri, ``Practical pulse-shaping
  waveforms for reduced-cyclic-prefix {OTFS},'' \emph{IEEE Transactions on
  Vehicular Technology}, vol.~68, no.~1, pp. 957--961, 2019.

\bibitem{Bellanger2010}
M.~Bellanger, ``{FBMC} physical layer: {A} primer,'' in \emph{{PHYDYAS} FP7
  Project Document}, Jan. 2010, pp. 1--31.

\bibitem{ShengliZhouBook}
S.~Zhou and Z.~Wang, ``Detection, synchronization and {Doppler} scale
  estimation,'' in \emph{{OFDM} for Underwater Acoustic Communications}.\hskip
  1em plus 0.5em minus 0.4em\relax Wiley, 2014, pp. 91--116.

\end{thebibliography}

\clearpage
\newpage
\section{Supplementary Material}
\subsection{Proof of \eqref{eqn:omega_conv}}{\label{supp:proof_eqn_omega_conv}}
We show that a cascade of two delay-scale channels, as shown in Figure \ref{fig:cascade_DS_channel}, is equivalent to a single channel, i.e., $\Pi_{h_2}(\Pi_{h_1}(s)) = \Pi_{h}(s)$, where $h(\tau,\alpha) = h_2(\tau,\alpha) \circledast_\omega  h_1(\tau,\alpha)$ and the symbol $\circledast_\omega$ denotes the $\omega$-convolution defined by
	\begin{equation}  \label{eqn:omega_conv_}
	h(\tau,\alpha) = \iint h_2(\tau',\alpha') h_1\left(\alpha'(\tau-\tau'),\frac{\alpha}{\alpha'}\right) d\tau' d\alpha'.
	\end{equation}	
	
	To show this, we substitute
	\begin{equation}
	v(t) = \iint h_1(\tau'',\alpha'')  \sqrt{\alpha''} u\left(\alpha''(t-\tau'')\right) d\tau'' d\alpha'', \nonumber
	\end{equation} 
	into the relation 
	\begin{equation}
	r(t) = \iint h_2(\tau',\alpha') \sqrt{\alpha'} v\left(\alpha'(t-\tau'),\right) d\tau' d\alpha', \nonumber
	\end{equation} 
	and make the change of variables, $\tau'' \rightarrow \alpha'(\tau - \tau')$ and $ \alpha'' \rightarrow \alpha/\alpha'$, to obtain
	\begin{align}  \label{eqn:omegaConv_}
	r(t) &=   \iint \iint  h_2(\tau',\alpha') h_1\left(\alpha'(\tau-\tau'),\frac{\alpha}{\alpha'}\right) d\tau' d\alpha' \nonumber \\ &\hspace{1.5cm} \times \sqrt{\alpha} u(\alpha(t-\tau)) d\tau d\alpha \nonumber \\
	&= \int\int h(\tau, \alpha) \sqrt{\alpha} u(\alpha (t - \tau)) d\tau d\alpha, \nonumber
	\end{align}	
	where $h(\tau, \alpha)$ is given by \eqref{eqn:omega_conv_}.

\subsection{Derivation of \eqref{eqn:odssTransform}}{\label{supp:proof_odssTransform}}
Using \eqref{eqn:sdSignal} and the definition of $\omega$-convolution in \eqref{eqn:omega_conv}, we can express ${f(\tau,\alpha) = h(\tau,\alpha) \circledast_\omega X(\tau,\alpha)}$ as
\begin{eqnarray}  \label{eqn:odssTransform_}
f(\tau,\alpha) &=& \iint h(\tau', \alpha') \sum_n \sum_m  X[n,m] \nonumber\\ 
& & \times \delta\left(\alpha'(\tau-\tau')-\textcolor{black}{\frac{m}{q^nW}},\frac{\alpha}{\alpha'}-q^n\right) d\tau' d\alpha'  \nonumber\\
&=&  \sum_n \sum_m X[n,m]  \iint h(\tau', \alpha') \nonumber\\ 
& & \times \delta\left(\alpha'(\tau-\tau')-\textcolor{black}{\frac{m}{q^nW}},\frac{\alpha-q^n \alpha'}{\alpha'}\right) d\tau' d\alpha' \nonumber\\
&\overset{(e)}{=}&  \sum_n \sum_m X[n,m]  h\left(\tau-\textcolor{black}{\frac{m}{\alpha W}}, \frac{\alpha}{q^n} \right) \textcolor{black}{q^{-n}}, \nonumber \\ 
\end{eqnarray}	
where the last equality (e) follows from the following properties of the Dirac delta function: 
\begin{enumerate}
	\item Generalized scaling property:	
	\begin{equation} 
	\delta(\mathbf{g}(\mathbf{x})) = \underset{{\lbrace{ \mathbf{x}_0: \mathbf{g}(\mathbf{x}_0)=\mathbf{0} \rbrace}}}{\sum} \frac{\delta(\mathbf{x}-\mathbf{x}_0)}{\left|{\text{det }\frac{\partial (g_1,\ldots,g_n)}{\partial (x_1,\ldots,x_n)}}\right|_{\mathbf{x} = \mathbf{x}_0}}, \nonumber
	\end{equation}
	where $\mathbf{g}: \mathbb{R}^n \rightarrow \mathbb{R}^n$ is a bi-Lipschitz function and $\left|\text{det }\frac{\partial (g_1,\ldots,g_n)}{\partial (x_1,\ldots,x_n)}\right| \ne 0$ for $\mathbf{x} \in \lbrace{\mathbf{x}_0: \mathbf{g}(\mathbf{x}_0)=\mathbf{0} \rbrace}$.
	
	\item Sifting property:
	
	\begin{equation}
	\int  \mathbf{f}(\mathbf{x}) \delta(\mathbf{x}-\mathbf{x}_0) d\mathbf{x} = \mathbf{f}(\mathbf{x}_0). \nonumber
	\end{equation}
\end{enumerate}

\subsection{Derivation of \eqref{eqn:odssIo}}{\label{supp:proof_odssIO}}
	
	\begin{eqnarray}  \label{eqn:odssDemod_}
	A_{g_{\text{rx}},r_s}(\tau,\alpha) &=& \int g^*_{\text{rx}}\left(\alpha(t-\tau)\right)  \sqrt{\alpha} r_s(t) dt \nonumber\\
	&=& \int g^*_{\text{rx}}\left(\alpha(t-\tau)\right) \sqrt{\alpha} \iint f(\tau',\alpha') \nonumber \\
	& & \times  \sqrt{\alpha'} g_{\text{tx}}\left(\alpha'(t-\tau')\right) d\tau' d\alpha'  dt \nonumber \\
	&=&  \iint f(\tau',\alpha') \int g^*_{\text{rx}}\left(\alpha(t-\tau)\right) \nonumber \\
	& & \times \sqrt{\alpha \alpha'}  g_{\text{tx}}\left(\alpha'(t-\tau')\right) dt d\tau' d\alpha' \nonumber \\
	&=&  \iint f(\tau',\alpha') \nonumber \\
	& & \times \int g^*_{\text{rx}}\left(\frac{\alpha}{\alpha'}\left(t'-\alpha'\left(\tau-\tau'\right)\right)\right) \nonumber \\
	& & \times \sqrt{\frac{ \alpha}{ \alpha'}} g_{\text{tx}}(t') dt' d\tau' d\alpha' \nonumber \\
	&=&  \iint f(\tau',\alpha') \nonumber \\
	& & \times A_{g_{\text{rx}},g_{\text{tx}}}\left(\alpha'(\tau-\tau'),\frac{\alpha}{\alpha'}\right) d\tau' d\alpha' \nonumber \\
	&=& f(\tau,\alpha) \circledast_\omega A_{g_{\text{rx}},g_{\text{tx}}}(\tau, \alpha).
	\end{eqnarray}	
	
	It is now straightforward to see that
	\begin{eqnarray}  \label{eqn:supp_odssIo_}
	A_{g_{\text{rx}},r_s}(\tau,\alpha)  &=&  f(\tau,\alpha) \circledast_\omega A_{g_{\text{rx}},g_{\text{tx}}}(\tau, \alpha) \nonumber \\
	&\overset{(e)}{=}& \iint  \sum_n \sum_m X[n,m]  \nonumber \\ 
	& & \times h\left(\tau'-\frac{m}{\alpha' W}, \frac{\alpha'}{q^n} \right) q^{-n}  \nonumber \\
	& & \times A_{g_{\text{rx}},g_{\text{tx}}}\left(\alpha'(\tau-\tau'),\frac{\alpha}{\alpha'}\right) d\tau' d\alpha' \nonumber \\
	&\overset{(f)}{=}& \sum_n \sum_m X[n,m] H_{n,m}(\tau, \alpha),
	\end{eqnarray}	
	where we used \eqref{eqn:odssTransform} at step (e), and the following definition at step (f):
	\begin{multline}  
	H_{n,m}(\tau, \alpha) \triangleq \iint h\left(\tau'-\frac{m}{\alpha' W}, \frac{\alpha'}{q^n} \right) q^{-n}   \\
	\times A_{g_{\text{rx}},g_{\text{tx}}} \left( \alpha'(\tau-\tau'), \frac{\alpha}{\alpha'} \right) d\tau' d\alpha'. \label{eqn:Hnm_DS0}
	\end{multline}
	By a transformation of variables, $\tau'' = \tau'-\frac{m}{\alpha' W}$ and $\alpha'' = \frac{\alpha'}{q^n}$, and using the fact that the determinant of the Jacobian of this transformation is $q^n$, we can rewrite the double integral in \eqref{eqn:Hnm_DS0} as
	\begin{multline}  
	H_{n,m}(\tau, \alpha) = \iint  h(\tau'',\alpha'')  \\ 
	\times A_{g_{\text{rx}},g_{\text{tx}}}\left(\alpha'' q^n \left( \tau-\frac{m}{\alpha'' q^n W}-\tau''\right), \frac{\alpha}{\alpha''q^n}\right) d\tau'' d\alpha''. \label{eqn:supp_Hnm_DS}
	\end{multline}

\subsection{{Pulse Shaping Functions for OTFS}}{\label{supp:proof_OTFS_NonBiorthogonality}}

The narrowband cross-ambiguity function, apropos of OTFS modulation, between the transmit and receive pulse shaping functions is defined by (see \eqref{eqn:xAmbiguity})
\begin{equation}  \label{eqn:Tx_Rx_xAmbiguity}
A_{g_{\text{rx}},g_{\text{tx}}}(\tau,\nu) \triangleq \int_t e^{-j2\pi\nu(t-\tau)} g^*_{\text{rx}}(t-\tau)  g_{\text{tx}}(t) dt.
\end{equation}

It is clear from \eqref{eqn:Tx_Rx_xAmbiguity} that, for a given $\tau$, the functions $\Psi(\nu) = A_{g_{\text{rx}},g_{\text{tx}}}(\tau,\nu)e^{-j2\pi\nu\tau}$ and $\psi(t) = g^*_{\text{rx}}(t-\tau)  g_{\text{tx}}(t)$ are Fourier pairs, and therefore
\begin{equation}  \label{eqn:Tx_Rx_xAmbiguity_IFT}
g^*_{\text{rx}}(t-\tau)  g_{\text{tx}}(t)  = \int_{\nu} A_{g_{\text{rx}},g_{\text{tx}}}(\tau,\nu) e^{j2\pi\nu(t-\tau)}  d\nu.
\end{equation}


Consider, for example,  a narrowband cross-ambiguity function, $A_{g_{\text{rx}},g_{\text{tx}}}(\tau,\nu)$, which is non-zero only on $S =  I_{\tau}  \times  I_{\nu}$, where $ I_{\tau} = \left( - \tau_{\max}, \tau_{\max} \right)$ and $ I_{\nu} =\left(- \nu_{\max}, \nu_{\max} \right) $, and $A_{g_{\text{rx}},g_{\text{tx}}}(\tau,\nu) = 1, \forall (\tau, \nu) \in S$. In this case, we see from \eqref{eqn:Tx_Rx_xAmbiguity_IFT} that
\begin{equation}  \label{eqn:Tx_Rx_pulse_relation}
g^*_{\text{rx}}(t-\tau)  g_{\text{tx}}(t)  = \begin{cases}
2 \nu_{\max} \text{ sinc}\left(2\nu_{\max}(t-\tau)\right), & \tau \in I_{\tau} \\
0, & \text{otherwise}.
\end{cases}
\end{equation}

This is clearly impossible: to have $g^*_{\text{rx}}(t-\tau)  g_{\text{tx}}(t)$ and its Fourier transform ($t \rightarrow \nu$) $A_{g_{\text{rx}},g_{\text{tx}}}(\tau,\nu)$ to be both finitely supported. Hence, we cannot design transmit and receive filters that exactly satisfy robust bi-orthogonality. Also, more generally, even if $A_{g_{\text{rx}},g_{\text{tx}}}(\tau,\nu)$ is not identically unity $\forall (\tau, \nu) \in S$, due to Heisenberg's uncertainty principle, it is still impossible to find transmit and receive filters that are robustly bi-orthogonal~\cite{Raviteja2018}.

\subsection{{Pulse Shaping Functions for ODSS}}{\label{supp:proof_ODSS_NonBiorthogonality}}

In ODSS, the wideband cross-ambiguity function between the transmit and receive pulse shaping functions is defined by (see \eqref{eqn:Xambiguity})
\begin{equation}  \label{eqn:Tx_Rx_Xambiguity}
A_{g_{\text{rx}},g_{\text{tx}}}(\tau,\alpha) \triangleq \int g^*_{\text{rx}}\left(\alpha(t-\tau)\right)  \sqrt{\alpha} g_{\text{tx}}(t)  dt.
\end{equation}
Let $G_{\text{tx}}(f)$ and $G_{\text{rx}}(f)$ denote the Fourier transforms of the transmit and receive pulse shaping functions, $g_{\text{tx}}(t)$ and $g_{\text{rx}}(t)$, respectively. Then, by Parseval's theorem, we can express the integral in \eqref{eqn:Tx_Rx_Xambiguity} as
\begin{equation}
A_{g_{\text{rx}},g_{\text{tx}}}(\tau,\alpha)  = \frac{1}{\sqrt{\alpha}} \int G^*_{\text{rx}}\left(\frac{f}{\alpha}\right) G_{\text{tx}}(f)  e^{j2 \pi f \tau} df. \label{eqn:FD_Xambiguity}
\end{equation}

It is clear from \eqref{eqn:FD_Xambiguity} that, for a given $\alpha$, the functions $\psi(\tau) = A_{g_{\text{rx}},g_{\text{tx}}}(\tau,\alpha)$ and $\Psi(f) = \frac{1}{\sqrt{\alpha}} G^*_{\text{rx}}\left(\frac{f}{\alpha}\right) G_{\text{tx}}(f)$ are Fourier pairs, and therefore
\begin{equation}  \label{eqn:Tx_Rx_Xambiguity_IFT}
\frac{1}{\sqrt{\alpha}} G^*_{\text{rx}}\left(\frac{f}{\alpha}\right) G_{\text{tx}}(f)  = \int A_{g_{\text{rx}},g_{\text{tx}}}(\tau,\alpha) e^{-j2 \pi f \tau}  d\tau.
\end{equation}

Setting $f = \alpha$ in \eqref{eqn:Tx_Rx_Xambiguity_IFT}, we obtain
\begin{equation}  \label{eqn:Tx_Rx_Xambiguity_IFT_t_eq_tau}
\frac{1}{\sqrt{\alpha}} G^*_{\text{rx}}(1)  G_{\text{tx}}(\alpha)  = \int A_{g_{\text{rx}},g_{\text{tx}}}(\tau,\alpha) e^{-j2 \pi \alpha \tau}  d\tau,
\end{equation}
from where, by replacing $\alpha$ with $f$, we can determine $G_{\text{tx}}(f)$ up to a scale factor $1/G^*_{\text{rx}}(1)$, provided $G^*_{\text{rx}}(1) \ne 0$. On the other hand, setting $f = 1$ and replacing $\alpha$ with $1/\alpha$ in \eqref{eqn:Tx_Rx_Xambiguity_IFT}, leads to
\begin{equation}  \label{eqn:Tx_Rx_Xambiguity_IFT_t_eq_0}
\sqrt{\alpha} G^*_{\text{rx}}(\alpha)  G_{\text{tx}}(1)= \int A_{g_{\text{rx}},g_{\text{tx}}}(\tau,1/\alpha) e^{-j 2 \pi \tau} d\tau,
\end{equation}
from where, by replacing $\alpha$ with $f$, we can find $G_{\text{rx}}(f)$ up to a scale factor $1/G^*_{\text{tx}}(1)$, provided $G^*_{\text{tx}}(1)\ne 0$. We determine the transmit and receive pulse shaping functions in the time domain by Fourier inversion of $G_{\text{tx}}(f)$ and $G_{\text{rx}}(f)$, respectively.

Consider, for example, a wideband cross-ambiguity function, $A_{g_{\text{rx}},g_{\text{tx}}}(\tau,\alpha)$, which is non-zero only for $S = \lbrace (\tau, \alpha): \tau \in \left( - \tau_{\max}, \tau_{\max} \right), \alpha \in \left(1/ \alpha_{\max}, \alpha_{\max} \right) \rbrace$, where $\alpha_{\max}  \ge 1 $, and $A_{g_{\text{rx}},g_{\text{tx}}}(\tau,\alpha) = e^{j 2 \pi f_0 \tau}/\sqrt{\alpha}$, for $(\tau, \alpha) \in S$, where $f_0 = \frac{\alpha^{-1}_{\max} + \alpha_{\max}}{2}$. In this case, we find that the Fourier transform of the transmit pulse shaping function is given by
\begin{equation}  \label{eqn:ODSS_Tx_Pulse}
G_{\text{tx}}(f)  = \begin{cases}
\frac{2 \tau_{\max}}{G^*_{\text{rx}}(1)} \text{sinc}\left((f-f_0)\tau_{\max}\right), & f \in  \left(\frac{1}{\alpha_{\max}}, \alpha_{\max}\right), \\
0, & \text{otherwise},
\end{cases}
\end{equation}
and the receive pulse shaping function in the Fourier domain is given by
\begin{equation}  \label{eqn:ODSS_Rx_Pulse}
G^*_{\text{rx}}(f)  = \begin{cases}
\frac{2 \tau_{\max}}{G_{\text{tx}}(1)} \text{sinc}\left((f_0-1)\tau_{\max}\right),  & f \in  \left(\frac{1}{\alpha_{\max}}, \alpha_{\max}\right), \\
0, & \text{otherwise}.
\end{cases}
\end{equation}

We immediately recognize the impossibility of designing \emph{finite} duration transmitter and receiver pulse shaping filters for ODSS modulation that satisfy the robust bi-orthogonality condition exactly. While the receiver pulse shaping filter, given by \eqref{eqn:ODSS_Rx_Pulse}, is exactly rectangular (in frequency domain), the transmitter pulse shaping filter, given by \eqref{eqn:ODSS_Tx_Pulse},  is nearly rectangular (in frequency domain) since for typical values of channel delay spread, $\tau_{\max}$, 
and Doppler scale, $\alpha_{\max}$, 
$\alpha_{\max}-1/\alpha_{\max} \ll \frac{1}{\tau_{\max}}$ holds. Therefore, the pulse shaping filters in the time domain cannot be finite duration waveforms. Conversely, finite duration pulse shaping filters cannot have spectra, as in \eqref{eqn:ODSS_Tx_Pulse} and \eqref{eqn:ODSS_Rx_Pulse}, and hence do not satisfy the robust bi-orthogonality condition exactly. We are thus left to choose filters that are nearly bi-orthogonal in a robust manner. 

\subsection{ODSS: Spectral Efficiency and Orthogonality}{\label{supp:specEff}}

\begin{figure}[h]
	\setxysizeo
	\centering
	\includegraphics[scale=.42]{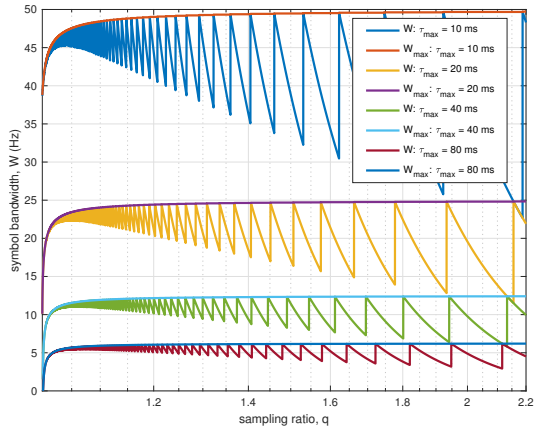}
	\caption{Plots of the transmit filter bandwidth, $W$  in \eqref{eqn:txFiltBW_q}, and maximum allowed bandwidth, $W_{\max}$ in \eqref{eqn:Wmax}, as the sampling ratio, $q$, is varied for various channel delay spreads. Small values of the transmit filter bandwidth, and hence a long symbol duration, needs to be used in channels with large delay spread.}	\label{fig:WWmaxvsQ}
	\pullUp
	\vspace{-0.1cm}
\end{figure}

\begin{figure}[h]
	\setxysizeo
	\centering
	\includegraphics[scale=.42]{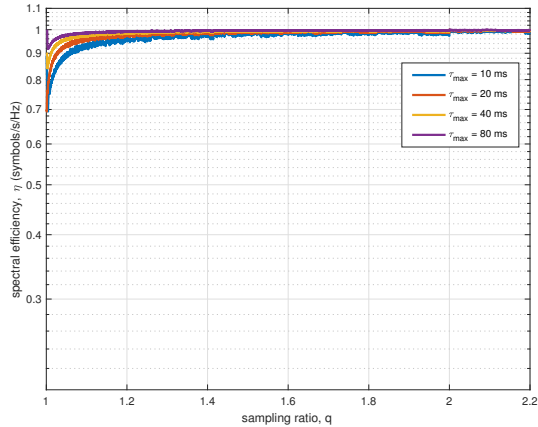}
	\caption{Plots of the spectral efficiency, $\eta$ (in symbols/s/Hz), of the ICI-free ODSS scheme as the sampling ratio, $q$, is varied. The spectral efficiency of the ODSS is quite close to unity for $q>1.2$. }\label{fig:EtavsQ}
	\pullUp
	\vspace{-0.1cm}			
\end{figure}

Consider the choice of ODSS modulator parameters for \textcolor{black}{an underwater} communication system of bandwidth $B = 10$ kHz operated in a channel with a Doppler spread of $\alpha_{\max} = 1.001$~\textcolor{black}{\cite{ShengliZhouBook}}. In Fig. \ref{fig:WWmaxvsQ}, we plot the transmit filter bandwidth, in \eqref{eqn:txFiltBW_q}, and the maximum allowed  bandwidth, given by \eqref{eqn:Wmax}, for different channel delay spreads as the sampling ratio $q$ is varied. 
The spectral efficiency of the ODSS modulation with these parameters and $\gamma = 2$ is plotted as a function of $q$  in Fig. \ref{fig:EtavsQ}.

\begin{figure}[h]
	\setxysizeo
	\centering
	\includegraphics[scale=.42]{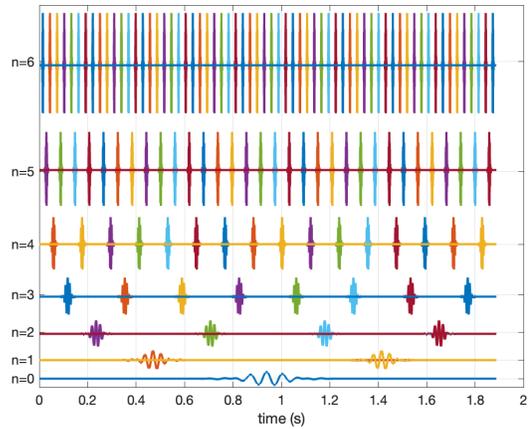}
	\caption{\textcolor{black}{ODSS subcarriers, for $n=0,1,\ldots,6$, on a dyadic ($q=2$) tiling over an ODSS symbol  duration of $T=1.9$ seconds. Note that a total of $N_7 = 127$ subcarriers are tiled in the symbol duration. }}	\label{fig:odssSubcarrierWaveforms}
	\pullUp
	\vspace{-0.1cm}
\end{figure}

Fig. \ref{fig:odssSubcarrierWaveforms} shows the ODSS subcarrier waveforms, constructed as discussed in Sec. \ref{sec:pulseShapeFn}, on a dyadic ($q=2$) tiling in the delay-scale space for a symbol block duration of $T=1.9$ seconds and time-scale indices $n=0,1,\ldots,6$.

\begin{figure}[h]
	\setxysizeo
	\centering
	\includegraphics[scale=.42]{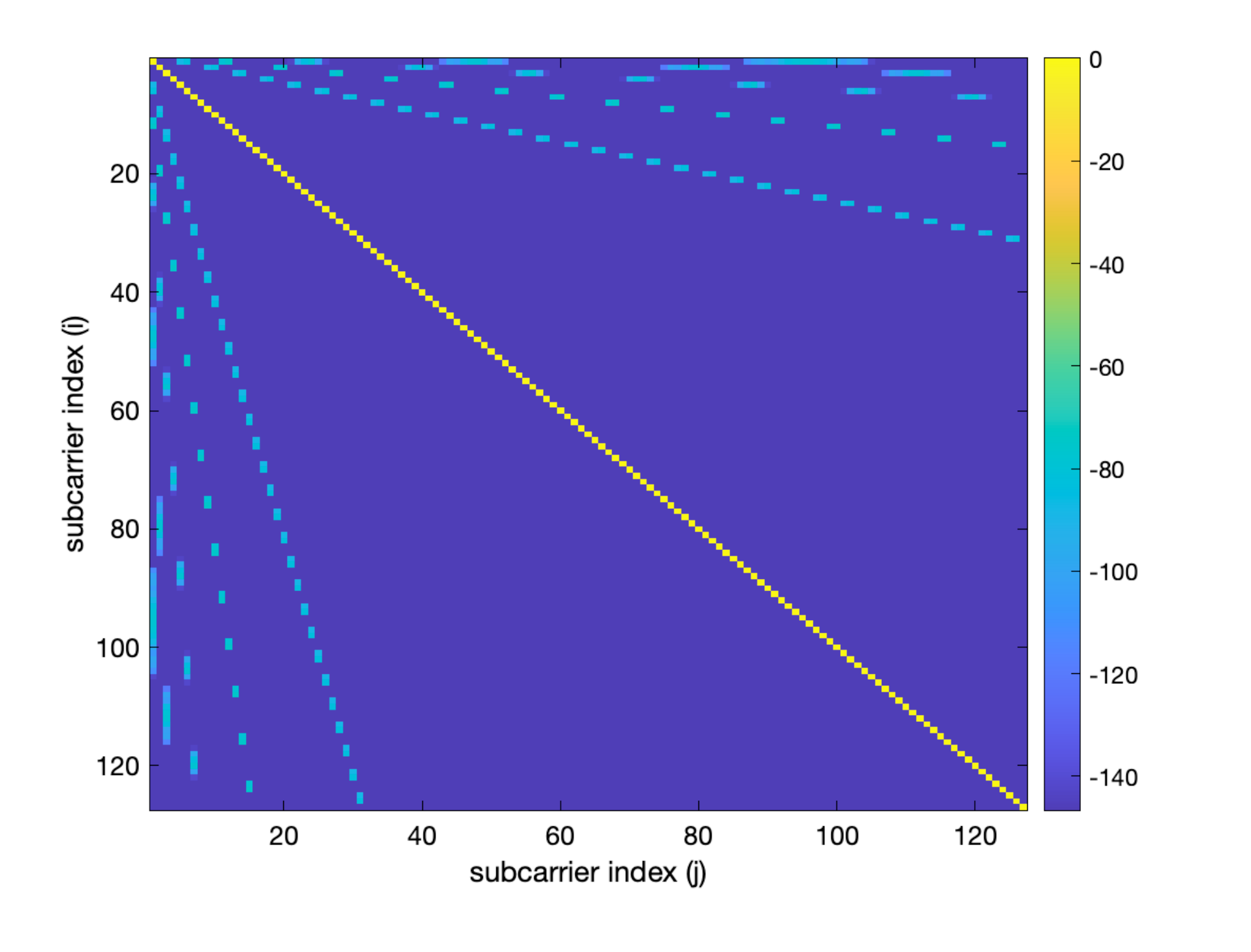}
	\caption{\textcolor{black}{ODSS subcarrier waveform correlation matrix, for $n=0,1,\ldots,6$. The normalized correlation (in dB) values are color coded and displayed. The peak intensity corresponds to $0$ dB (yellow). The  cross-correlation between any two distinct ODSS subcarrier waveforms is less than $-74$ dB.}}	\label{fig:odssSubcarrierCorrelation}
	\pullUp
	\vspace{-0.1cm}
\end{figure}

\textcolor{black}{The pairwise correlations of the ODSS subcarrier waveforms are shown as an image in Fig. \ref{fig:odssSubcarrierCorrelation}: the intensity of $(m,n)$th cell denotes the magnitude of the correlation between the $m$th and $n$th subcarrier waveforms. We see that 
	the correlation matrix is nearly diagonal as the normalized cross-correlation between any two distinct ODSS subcarrier waveforms is less than $-74$ dB.}

\newpage
\subsection{ODSS Channel Matrix}{\label{supp:plots}}

\begin{figure}[h]
	\setxysizeo
	\centering
	\includegraphics[scale=.42]{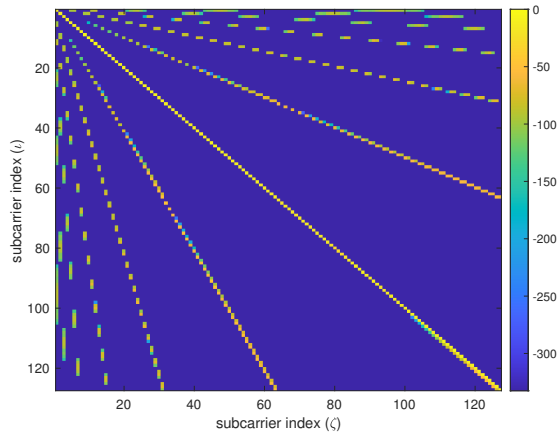}
	\caption{\textcolor{black}{ODSS channel matrix (normalized magnitude, in dB) is nearly diagonal in the delay-scale domain with the maximum ICI level not exceeding $-25$ dB. Channel equalization can be implemented by multiplying each delay-scale domain measurement with a respective complex number.}}	\label{fig:odssDelayScaleDomainChanMat}
	\pullUp
	\vspace{-0.1cm}
\end{figure}

\begin{figure}[h]
	\setxysizeo
	\centering
	\includegraphics[scale=.42]{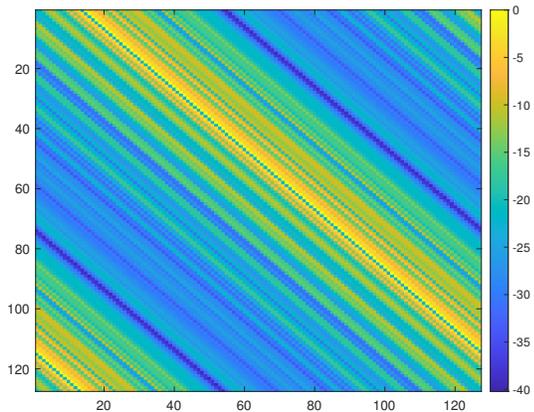}
	\caption{\textcolor{black}{ODSS channel matrix (normalized magnitude, in dB) in the Mellin-Fourier domain. Channel equalizer complexity in the Mellin-Fourier domain is high due to the non-sparse nature of the associated channel matrix.}}	\label{fig:odssMellinFourierDomainChanMat}
	\pullUp
	\vspace{-0.1cm}
\end{figure}

\textcolor{black}{Figures \ref{fig:odssDelayScaleDomainChanMat} and \ref{fig:odssMellinFourierDomainChanMat} show the ODSS channel matrix in the delay-scale and Mellin-Fourier domains, respectively, for one of the simulated channel instances. The ODSS channel matrix in the delay-scale domain is nearly diagonal and, therefore, a single-tap channel equalizer can be implemented in this domain.}

\newpage
\subsection{Effect of Time-scaling: ODSS versus OFDM}{\label{supp:ICI_ofdm_otfs}}

\begin{figure}[h]
	\setxysizeo
	\centering
	\includegraphics[scale=.42]{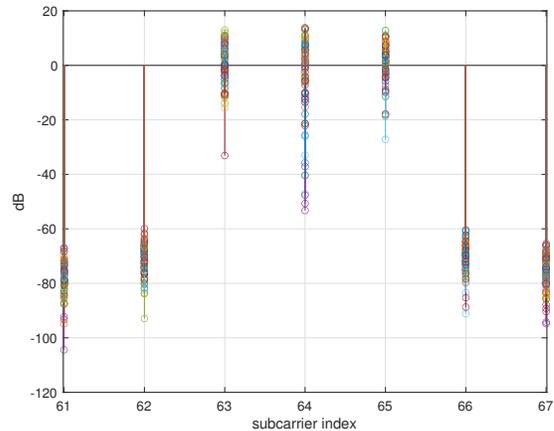}
	\caption{OFDM receiver processed subcarrier outputs. Only the $64$th subcarrier is transmitted with a BPSK symbol across a channel with $\alpha_{\max} = 1.001$. A Doppler shift of $\delta f_c = (\alpha_{\max} -1) f_ c = 12.8$ Hz ( $> \Delta F$) is experienced by the 64th subcarrier (corresponding to $f_c = 12.8$ kHz). Spurious pickups due to ICI can be observed on the $63$rd and $65$th subcarriers.}\label{fig:ofdmICI}
	\pullUp
	\vspace{-0.2cm}
	\end{figure}

\begin{figure}[h]
	\setxysizeo
	\centering
	\includegraphics[scale=.42]{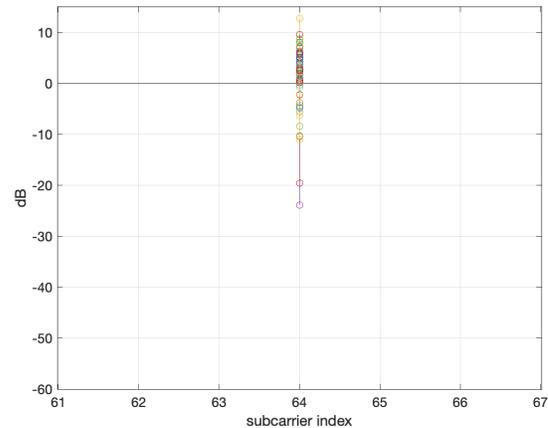}
	\caption{ODSS receiver processed subcarrier outputs in a delay-scale spread channel. Only the $64$th subcarrier is transmitted with a BPSK symbol across a channel with $\alpha_{\max} = 1.001$. Unlike OFDM, ICI is avoided in ODSS. 
	}\label{fig:odssICI}
	\pullUp
	\vspace{-0.2cm}
	\end{figure}

\textcolor{black}{To illustrate the effect of Doppler on OFDM and ODSS, we mount a symbol only on one of the subcarriers, say, the $n_{\text{sub}}=64$. We then observe the processed subcarrier outputs, for $100$ channel realizations, in the neighborhood of $n_{\text{sub}}$. In Figs. \ref{fig:ofdmICI} and \ref{fig:odssICI}, we plot the OFDM and ODSS receiver outputs, respectively. Spurious pickup by the subcarriers $n_{-}=63$ and $n_{+}=65$ in OFDM is due to Doppler. In  contrast, the ICI due to time-scaling effect of the channel is negligible in ODSS, thus illustrating its suitability for such channels.}

\begin{figure}[h]
	\setxysizeo
	\centering
	\includegraphics[scale=.42]{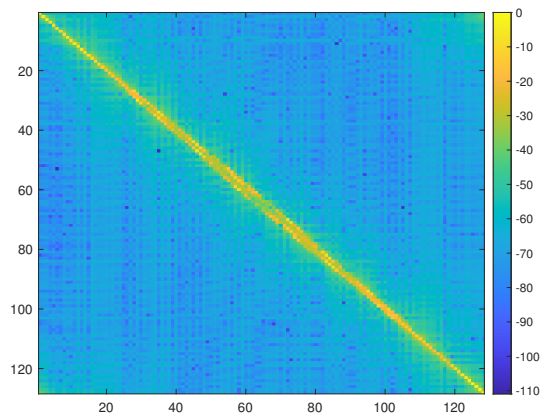}
	\caption{\textcolor{black}{OTFS channel matrix (normalized magnitude, in dB) in the time-frequency domain for a simulated delay-scale channel realization. The channel matrix is non-diagonal with a severe ICI in excess of $0$ dB relative to some diagonal entries. }}	\label{fig:otfsTfDomainChanMat}
	\pullUp
	\vspace{-0.1cm}
\end{figure}

\end{document}